\documentclass[superscriptaddress,
twocolumn,amsmath,amssymb]{revtex4}
\usepackage[dvips]{graphicx}% Include figure files
\usepackage{epsf}
\usepackage{epsfig}
\usepackage{latexsym}
\usepackage{amssymb}
\usepackage{amsfonts,amsbsy}
\usepackage{amsmath}
\usepackage{dcolumn}% Align table columns on decimal point
\usepackage{bm}
\usepackage{pifont}
\begin{document}
\title{Intra-cellular transport by single-headed kinesin KIF1A: 
effects of single-motor mechano-chemistry and steric interactions}
% Force line breaks with \\
\author{Philip Greulich}%
% \email{pg@thp.uni-koeln.de}
\affiliation{%
Institut  f\"ur Theoretische  Physik, Universit\"at
zu K\"oln D-50937 K\"oln, Germany
}%
\author{Ashok Garai}
%\email{debch@iitk.ac.in} 
\affiliation{%
Department of Physics, Indian Institute of Technology,
Kanpur 208016, India.
}%
\author{Katsuhiro Nishinari}
% \email{tknishi@mail.ecc.u-tokyo.ac.jp}
\affiliation{%
Department of Aeronautics and Astronautics,
Faculty of Engineering, University of Tokyo,
Hongo, Bunkyo-ku, Tokyo 113-8656, Japan.
}%
\author{Andreas Schadschneider}%
% \email{as@thp.uni-koeln.de}
\affiliation{%
Institut  f\"ur Theoretische  Physik, Universit\"at 
zu K\"oln D-50937 K\"oln, Germany
}%
\affiliation{%
Interdisziplin\"ares Zentrum f\"ur komplexe Systeme,
University of Bonn, Germany}%
\author{Debashish Chowdhury}%
%\email{debch@iitk.ac.in} 
\affiliation{%
Department of Physics, Indian Institute of Technology,
Kanpur 208016, India.
}%
\affiliation{%
Max-Planck Institute for Physics of Complex Systems,
N\"othnitzer Strasse 38, D-01187 Dresden, Germany.  
}%
\date{\today}% It is always \today, today,
\begin{abstract}
  In eukaryotic cells, many motor proteins can move simultaneously on
  a single microtubule track. This leads to interesting collective
  phenomena like jamming. Recently we reported ({\it Phys. Rev. Lett.
    {\bf 95}, 118101 (2005)}) a lattice-gas model which describes
  traffic of unconventional (single-headed) kinesins KIF1A. Here we
  generalize this model, introducing a novel interaction parameter
  $c$, to account for an interesting mechano-chemical process which
  has not been considered in any earlier model. We have been able to
  extract all the parameters of the model, except $c$, from
  experimentally measured quantities. In contrast to earlier models of
  intra-cellular molecular motor traffic, our model assigns distinct
  ``chemical'' (or, conformational) states to each kinesin to account
  for the hydrolysis of ATP, the chemical fuel of the motor. Our model
  makes experimentally testable theoretical predictions. We determine
  the phase diagram of the model in planes spanned by experimentally
  controllable parameters, namely, the concentrations of kinesins and
  ATP. Furthermore, the phase-separated regime is studied in some
  detail using analytical methods and simulations to determine e.g.\ 
  the position of shocks.  Comparison of our theoretical predictions
  with experimental results is expected to elucidate the nature of the
  mechano-chemical process captured by the parameter $c$.
\end{abstract}
%%%%%%%%%%%%%%%%%%%%%%%%%%%%%%%%%%%%%%%%%%%%%%%%%%%%%%%%%%%%%%%%%
\maketitle

%%%%%%%%%%%%%%%%%%%%%%%%%%%%%%%%%%%%%%%%%%%%%%%%%%%%
\section{\label{sec1}Introduction}
%%%%%%%%%%%%%%%%%%%%%%%%%%%%%%%%%%%%%%%%%%%%%%%%%%%%

Motor proteins are responsible for intra-cellular transport of wide 
varieties of cargo from one location to another in eukaryotic cells 
\cite{schliwa,molloy,jphys}. One crucial feature of these motors is 
that these move on filamentary tracks \cite{polrev}. Microtubules 
and filamentary actin are protein filaments which form part of a 
dual-purpose scaffolding called cytoskeleton \cite{howard}; these 
filamentary proteins act like struts or girders for the cellular 
architecture and, at the same time, also serve as tracks for the 
intra-cellular transportation networks. Kinesins and dyneins are two 
superfamilies of motors that move on microtubule whereas myosins move 
on actin filaments. A common feature of all these molecular motors is 
that these perform mechanical work by converting some other form of 
input energy. However, there are several crucial differences between 
these molecular motors and their macroscopic counterparts; the major 
differences arise from their negligibly small inertia. That's why the 
mechanisms of single molecular motors \cite{schliwa,molloy,jphys} and 
the details of the underlying mechano-chemistry \cite{bustamante} 
have been investigated extensively over the last two decades. 

However, often a single filamentary track is used simultaneously by 
many motors and, in such circumstances, the inter-motor interactions 
cannot be ignored. Fundamental understanding of these collective 
physical phenomena may also expose the causes of motor-related 
diseases (e.g., Alzheimer's disease) \cite{traffd} thereby helping, 
possibly, also in their control and cure. 

To our knowledge, the first attempt to understand effects of steric 
interactions of motors was made in the context of ribosome traffic 
on a single mRNA strand \cite{macdonald}. This led to the model 
which is now generally referred to as the totally asymmetric simple 
exclusion process (TASEP); this is one of the simplest models of 
non-equilibrium systems of interacting driven particles 
\cite{sz,derrida,schuetz}. 
In the TASEP a particle can hop forward to the next lattice site, 
with a probability $q$ per time step, if and only if the target site 
is empty; updating is done throughout either in parallel or in the
random-sequential manner.

Some of the most recent generic theoretical models of interacting 
cytoskeletal molecular motors \cite{lipo,frey1,santen,popkov} 
are appropriate extensions of TASEP. In those models the motor is
represented by a self-driven particle and the dynamics of the model 
is essentially an extension of that of the TASEP \cite{sz,schuetz} 
that includes Langmuir-like kinetics of attachment and detachment 
of the motors. Two different approaches have been suggested. 
In the approach followed by Parmeggiani, Franosch and Frey (PFF model) 
\cite{frey1,frey2}, attachment and detachment of the motors is modelled,  
effectively, as particle creation and annihilation, respectively, on the 
track; the diffusive motion of the motors in the surrounding fluid 
medium is not described explicitly. In contrast, in the alternative 
formulation suggested by Lipowsky and co-workers \cite{lipo,lipo2}, 
the diffusion of motors in the cell is also modelled explicitly.

In reality, a motor protein is not a mere particle, but an enzyme 
whose mechanical movement is coupled with its biochemical cycle. 
In a recent letter \cite{nosc} we considered specifically the 
{\it single-headed} kinesin motor, KIF1A 
\cite{okada1,okada2,okada3,okada4,okada5}. 
The movement of a single KIF1A motor had already been modelled 
earlier \cite{okada2,sasaki} by a Brownian ratchet mechanism 
\cite{julicher,reimann}. In contrast to the earlier models 
\cite{lipo,frey1,santen,popkov} of molecular motor traffic, which take 
into account only the mutual interactions of the motors, our model 
explicitly incorporates also this Brownian ratchet mechanism of the 
individual KIF1A motors, including its biochemical cycle that involves 
{\it adenosine triphosphate (ATP) hydrolysis}.

The TASEP-like models predict the occurrence of shocks.
But since most of the bio-chemistry is captured in these models 
through a single effective hopping rate, it is difficult to make 
direct quantitative comparison with experimental data which depend 
on such chemical processes. In contrast, the model we proposed in 
ref.~\cite{nosc} incorporates the essential steps in the biochemical 
processes of KIF1A as well as their mutual interactions and involves 
parameters that have one-to-one correspondence with experimentally 
controllable quantities.

Here, we present not only more details of our earlier calculations 
but also many new results on the properties of single KIF1A motors 
as well as their collective spatio-temporal organization. Moreover, 
here we also generalize our model to account for a novel mechano-chemical 
process which has not received any attention so far in the literature. 
More specifically, two extreme limits of this generalized version of 
the model correspond to two different plausible scenarios of
ADP release by the motor enzymes. To our knowledge \cite{okadapc}, 
at present, it is not possible to rule out either of these two 
scenarios on the basis of the available empirical data. However, our 
new generalized model helps in prescribing clear quantitative 
indicators of these two mutually exclusive scenarios; use of these 
indicators in future experiments may help in identifying the true 
scenario. 
 
An important feature of the collective spatio-temporal organization 
of motors is the occurance of a shock or domain wall, which is 
essentially the interface between the low-density and high-density 
regions. We focus on the dependence of the position of the domain 
wall on the experimentally controllable parameters of the model. 
Moreover, we make comparisons between our model in the low-density 
regime with some  earlier models of single motors. We also compare 
and contrast the basic features of the collective organization in 
our model with those observed in the earlier generic models of 
molecular motor traffic.

%%%%%%%%%%%%%%%%%%%%%%%%%%%%%%%%%%%%%%%%%%%%%%%%%%%%%%%%%%
\section{Discretized Brownian ratchet model for KIF1A: general formulation}
\begin{figure}[tb]
\begin{center}
\includegraphics[width=0.45\textwidth]{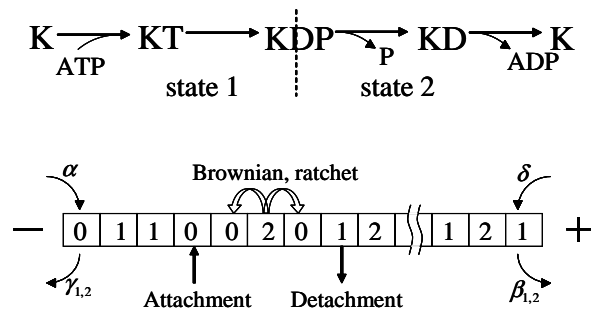}\\
\end{center}
\caption{A biochemical cycle of a KIF1A motor (upper) and
a three-valued discrete model for traffic of interacting
KIF1A motors on a finite microtubule filament (lower).
The states left to the dotted line in the upper figure
correspond to strongly bound to microtubule states (state 1)
while those right are weakly
bound (state 2).
0 denotes an empty site, and only 2 can move either to the
forward or backward site. Transition
from 1 to 2 occurs at the same site which corresponds hydrolysis,
and the detachment also happens in this process. The attachment is
possible only at the empty sites. At the minus and plus ends the
probabilities are different from those at sites in the bulk.}
\label{fig-lattice}
\end{figure}

%%%%%%%%%%%%%%%%%%%%%%%%%%%%%%%%%%%%%%%%%%%%%%%%%%%%%%%%%%
\label{sec2}
Through a series of in-vitro experiments, Okada, Hirokawa and
co-workers established \cite{okada1,okada2,okada3,okada4,okada5}
that:
\begin{enumerate}
{\renewcommand{\theenumi}{\roman{enumi}}
\item KIF1A molecule is an enzyme (catalyst) and in each enzymatic cycle it 
hydrolyzes one ATP molecule; the products of hydrolysis being adenosine 
diphosphate (ADP) and inorganic phosphate. Thus, each biochemical cycle of 
a KIF1A motor consists of four states: bare kinesin (K), kinesin bound with 
ATP (KT), kinesin bound with ADP and phosphate (KDP) and, finally, kinesin 
bound with only ADP (KD) after releasing phosphate (Fig.~\ref{fig-lattice}).
\item When a single-headed kinesin binds with a ATP molecule, its binding 
with its microtubule track is weakened by the ATP hydrolysis. Both K and KT 
bind strongly to microtubules. Hydrolysis of ATP leads to the state KDP 
which has a very short lifetime and soon yields KD by releasing phosphate.
KD binds weakly to a microtubule. After releasing all the products of 
hydrolysis (i.e., ADP and phosphate), the motor again binds strongly with 
the nearest binding site on the microtubule and thereby returns to the 
state K. 
\item In the state KD, the motor remains tethered to the microtubule 
filament by the electrostatic attraction between the positively charged 
$K$-loop of the motor and the negatively charged $E$-hook of the microtubule 
filament. Because of this tethering in the weakly bound state, a KIF1A 
cannot wander far away from the microtubule, but can execute (essentially 
one-dimensional) diffusive motion parallel to the microtubule filament. 
However, in the strongly bound state, the KIF1A motor cannot execute 
diffusive excursions away from the binding site on the microtubule.
}
\end{enumerate}

%%%%%%%%%%%%%%%%%%%%%%%%%%%%%%%%%%%%%%%%%%%%%%%
%%%%%%%%%%%%%%%%%%%%%%%%%%%%%%%%%%%%%%%%%%%%%%%

These experimental results for the biochemical cycle of KIF1A motors
indicate that a simplified description in terms of a 2-state model 
could be sufficient to understand the collective transport properties.
As shown in Fig.~\ref{fig-lattice} one distinguishes a state where
the motor is strongly bound to the microtubule (state 1) and 
a state where it is weakly bound (state 2).
It is worth pointing out that such a simplified 2-state model,
however, may not be adequate to capture the biochemical cycle
of other motors like, for example, conventional kinesins.
In such situations, a more detailed 4-state model is required.

As in the TASEP-type approach of the PFF model, the periodic array of 
the binding sites for KIF1A on the microtubule are represented as a 
one-dimensional lattice of sites that are labelled by the integer 
index $i$ ($i = 1,...,L$). KIF1A motors are represented by particles 
that can be in two different states 1 and 2, corresponding to the 
strongly-bound and weakly bound states. To account for the empirical 
observations, the model also contains elements of a Brownian ratchet. 
As in the PFF model, attachment and detachment of a motor are modelled 
as, effectively, creation and annihilation of the particles on the 
lattice. We use the random sequential update, and the dynamics of the 
system is given by the following rules of time evolution:\\

\noindent(1) {\it Bulk dynamics}\noindent\\
If the chosen site on the microtubule is empty, i.e., in state $0$,
then with probability $\omega_a dt$ a motor binds with the site causing a
transition of the state of the binding site from $0$ to $1$.
However, if the binding site is in state $1$, then it becomes $2$ with the
probability $\omega_h dt$ due to hydrolysis, or becomes $0$ with
probability $\omega_d dt$ due to the detachment from the microtubule during
hydrolysis.

If the chosen site is in state $2$, then the motor bound to this site 
steps forward to the next binding site in front by a ratchet mechanism 
with the rate $\omega_f$ or stays at the current location with the rate 
$\omega_s$. Both processes are triggered by the release of ADP. 
How should one modify these update rules if 
the next binding site in front is already occupied by another motor? 
Does the release of ADP from the motor, and its subsequent re-binding 
with the filamentary track, depend on the state of occupation of the next 
binding site in front of it? To our knowledge, experimental data available 
at present in the literature are inadequate to answer this question. 
Nevertheless, we can think of the two following plausible scenarios: 
in the cases $\cdots 2 1 \cdots$, or $\cdots 2 2 \cdots$,
the following kinesin, which is in state 2, can return to state 1,
only at its current location, with rate $\omega_s$ if ADP release
is regulated by the motor at the next site in front of it. But, if ADP 
release by the kinesin is independent of  the occupation status of the 
front site, then state 2 can return to state 1 at the fixed rate 
$\omega_s + \omega_f$, irrespective of whether or not the front site 
is occupied.

Therefore, we propose a generalization of our original model by 
incorporating both these possible scenarios within a single model
by introducing an interpolating parameter $c$ with $0\le c\le 1$. 
In this generalized version of our model, a motor in the state 2 
returns to the state 1 at the rate $\omega_s + (1-c)\omega_f$. 
The parameter $c$ $(0\le c\le 1)$ allows interpolation between the 
two above mentioned scenarios of ADP release by the kinesin.
For $c=1$ the transition from the strongly to the weakly bound state
in the ratchet mechanism depends on the occupation of the
front site. This is the case that has been treated in \cite{nosc}, 
where the release of ADP by a nucleotide-bound kinesin is tightly 
controlled by the kinesin at the next binding site in front of it. 
On the other hand, for $c<1$ the transition rate will depend partially 
on the occupation of the front site. For $c=0$ the ADP release process 
becomes completely independent of the state of the preceeding site.

As long as the motor does not release ADP, it executes random Brownian 
motion with the rate $\omega_b$.\\

\noindent(2) {\it Dynamics at the ends}\noindent\\
The probabilities of detachment and attachment at the two ends of the
microtubule can be different from those at any other site in the bulk.
We choose $\alpha$ and $\delta$, instead of $\omega_a$, as the
probabilities of attachment at the left and right ends. Similarly, we
take $\gamma_1$ and $\beta_1$, instead of $\omega_d$, as probabilities
of detachments at the left and right ends, respectively 
(Fig.~\ref{fig-lattice}). Finally, $\gamma_2$ and $\beta_2$, instead 
of $\omega_b$, are the probabilities of exit of the motors through the 
two ends by random Brownian movements.\\

\begin{figure}[htb]
\begin{center}
\includegraphics[angle=-90,width=0.4\textwidth]{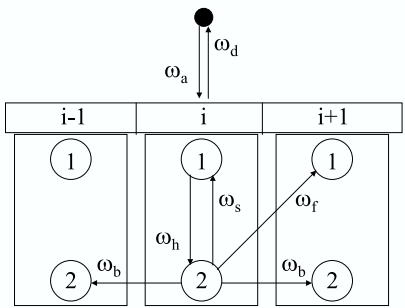}
\end{center}
\vspace{-1cm}
\caption{Schematic description of the three-state model of a 
single-headed kinesin motor that follows a Brownian ratchet 
mechanism. In the special case $2X\to 1X$, which has not been 
shown explicitly for the sake of simplicity, the rate constants 
would get modified following the prescriptions described in the text. 
}
\label{fig-kifbrat}
\end{figure}

For the dynamical evolution of the system, one of the $L$ sites is
picked up randomly and updated according to the rules given below
together with the corresponding probabilities (Fig.~\ref{fig-kifbrat}):
\begin{eqnarray}
 &&{\rm Attachment:} \,\,\,\,\,   0\to 1 \,\,\,{\rm with} \,\, \omega_a dt\\
 &&{\rm Detachment:} \,\,\,\,     1\to 0 \,\,\,{\rm with} \,\, \omega_d dt\\
 &&{\rm Hydrolysis:} \,\,\,\,\,\, 1\to 2 \,\,\,{\rm with} \,\, \omega_h dt\\
 &&{\rm Brownian\ motion:}\,\,\,\,\, \left\{\begin{array}{c}
     20 \to 02\,\,\,{\rm with} \,\, \omega_b dt\\
     02 \to 20\,\,\,{\rm with} \,\, \omega_b dt
   \end{array}\right.\\ \label{ratchet}
 &&{\rm Ratchet:}\,\,\,\, \left\{\begin{array}{l}
     20 \to 10\,\,\,\,\,{\rm with} \,\, \omega_s dt\\
     2X \to 1X\,\,\,{\rm with} \,\,
           \left(\omega_s + (1-c)\omega_f\right)dt\ \ \\
     20 \to 01\,\,\,\,\,{\rm with} \,\, \omega_f dt
   \end{array}\right.
\end{eqnarray}
Here $X$ denotes an occupied site irrespective of the chemical state 
of the motor, i.e., a site occupied by a motor that is in either state 
1 or state 2.

The ratchet mechanism (\ref{ratchet}) is triggered by the release 
of ADP and summarizes the transitions of a particle from state 2 to 
state 1. It distinguishes the two initial states $20$, where the front 
site is empty, and $2X$, where the front site is occupied. We see that 
the overall transition rate from state 2 to state 1 is 
$\omega_s+\omega_f$ if the front site is empty (initial state $20$), 
and it is $\omega_s+(1-c)\omega_f$ if the front site is occupied 
(initial state $2X$). This reflects the dependence of the ADP release 
rate on the front site occupation whenever $c\neq 0$.

%%%%%%%%%%%%%%%%%%%%%%%%%%%%%%%%%%%%%%%%%%%%%%%
%%%%%%%%%%%%%%%%%%%%%%%%%%%%%%%%%%%%%%%%%%%%%%%

\begin{figure}[ht]
\begin{center}
\includegraphics[width=0.9\columnwidth]{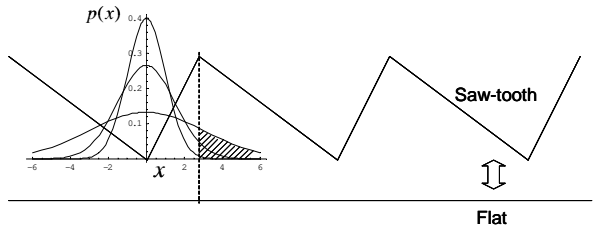}
\end{center}
\caption{The two forms of the time-dependent potential used for
implementing the Brownian ratchet mechanism.
}
\label{fig-ratgauss}
\end{figure}

The physical processes captured by the rate
constants $\omega_f$ and $\omega_s$ can be understood as follows by
analyzing the Brownian ratchet mechanism illustrated in
Fig.~\ref{fig-ratgauss}.
%%%%%%%%%%%%%%%%%%%%%%%%%%%%%%%%%%%%%%%%%%%%%%%%%%%%%%%%%%%%%%
%%%%%%%%%%%%%%%%%%%%%%%%%%%%%%%%%%%%%%%%%%%%%%%%%%%%%%%%%%%%%
For the sake of simplicity, we consider only one molecular motor, and
let us imagine
that the potential seen by the motor periodically oscillates between
the sawtooth shape and the flat shape shown in Fig.~\ref{fig-ratgauss}.
When the sawtooth form remains ``on'' for some time, the particle
settles at the bottom of a well. Then, when the potential is switched
``off'', the probability distribution of the position of the particle
is given by a delta function which, because of free diffusion in the
absence of any force, begins to spread. After some time the Gaussian
profile spreads to such an extent that it has some overlap also with
the well in front, in addition to the overlap it has with the original
well. At that stage, when the sawtooth potential is again switched on,
there is a non-vanishing probability that the particle will find
itself in the well in front; this probability is proportional to the
area of the hatched part of the Gaussian profile shown in
Fig.~\ref{fig-ratgauss} and is accounted for in our model by the
parameter $\omega_f$. There is also significant probability that the
particle will fall back into the original well; this is captured in
our model by the parameter $\omega_s$.

%%%%%%%%%%%%%%%%%%%%%%%%%%%%%%%%%%%%%%%%%%%%%%%%%%%%%%
\subsection{Parameters of the model}
%%%%%%%%%%%%%%%%%%%%%%%%%%%%%%%%%%%%%%%%%%%%%%%%%%%%%%

From experimental data \cite{okada1,okada3}, good estimates for the
parameters of the suggested model can be obtained.
The detachment rate $\omega_d \simeq 0.1$ s$^{-1}$ is found to
be independent of the kinesin population. On the other hand,
$\omega_a = 10^7$~$C$/M$\cdot$s depends on the concentration $C$
(in M) of the kinesin motors. In typical eucaryotic cells
{\it in-vivo} the kinesin concentration can vary between $10$ and
$1000$ nM. Therefore, the allowed range of $\omega_a$ is
$0.1$ s$^{-1} \leq \omega_a \leq 10$ s$^{-1}$.

\begin{figure}[tb]
\begin{center}
\includegraphics[width=0.4\textwidth]{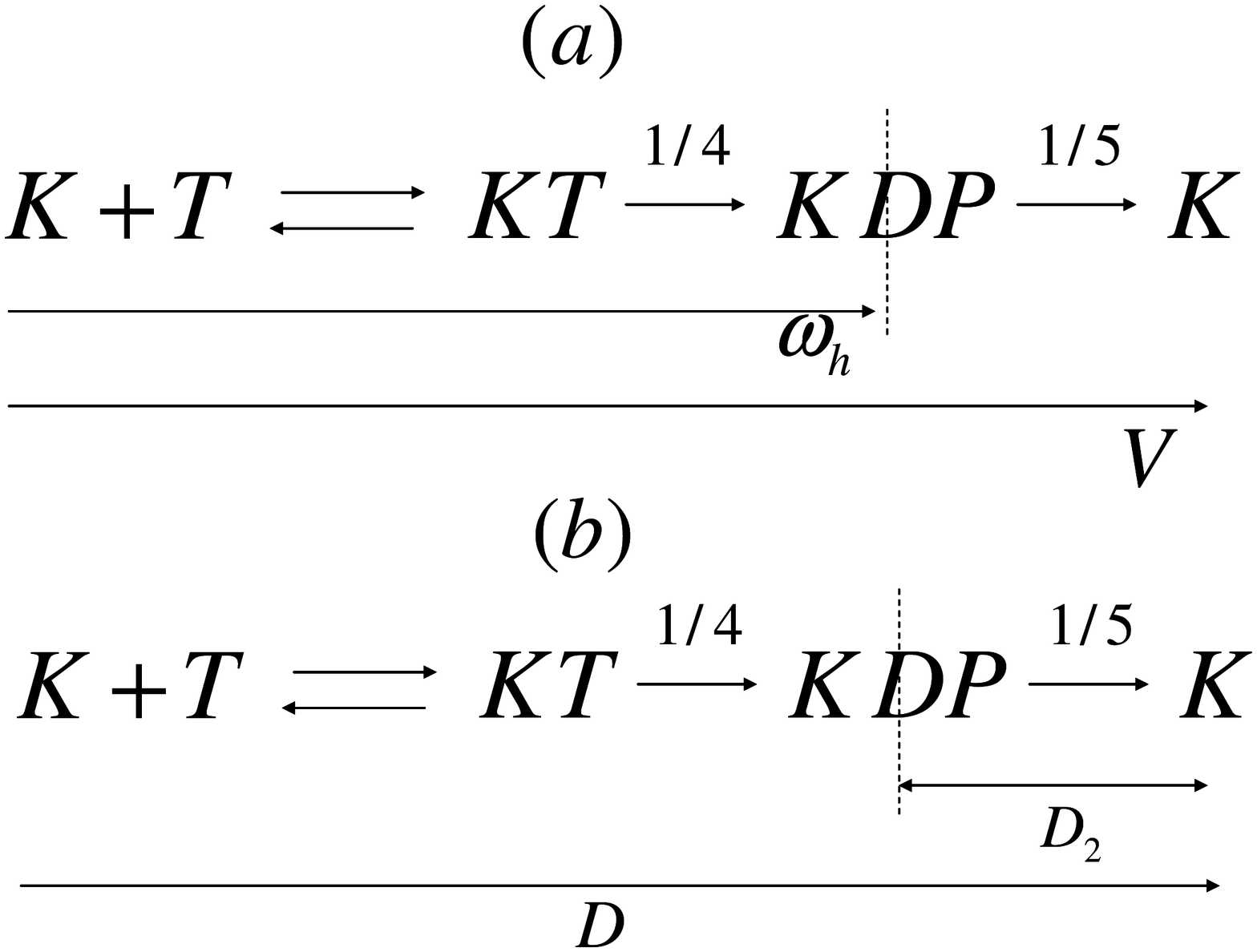}\\
\end{center}
\caption{The biochemical cycle of KIF1A is shown to define some 
important parameters which can be extracted from experimental 
data. See text for more details. 
}
\label{fig-exptpar}
\end{figure}

Total time taken for the hydrolysis of one ATP molecule is about
$9$~ms of which $4$~ms is spent in the state $1$ and $5$~ms in the
state $2$. The corresponding rates $1/4$ and $1/5$ are shown in
Fig.~\ref{fig-exptpar}. The motion of KIF1A is purely diffusive 
only when it is in the state $2$ and the corresponding diffusion 
coefficient is denoted by the symbol $D_2$. Using the measured 
diffusion constant $D = 40,000$~nm$^2$/s \cite{okada2} and the 
relation $D_2 = (9/5) D$, we get $D_2 = 72,000$~nm$^2$/s (see
Fig.~\ref{fig-exptpar}(b)). The time $\omega_b^{-1}$ must be such 
that $\omega_b \sim D_2/(8\text{nm})^2$, and, hence, we get 
$\omega_b \simeq 1125$~s$^{-1}$.

Moreover, from the experimental observations that the mean step
size is $3$~nm whereas the separation between the successive binding
sites on a microtubule is $8$~nm, we conclude $\omega_f/\omega_s \simeq 3/8$.
Furthermore, from the measured total time of each cycle, we estimate
that $\omega_s + \omega_f \simeq 200$~s$^{-1}$. From these two
relations between $\omega_f$ and $\omega_s$ we get the individual
estimates $\omega_s \simeq 145$~s$^{-1}$ and
$\omega_f \simeq 55$~s$^{-1}$.

Assuming the validity of the Michaelis-Menten type kinetics for the
hydrolysis of ATP \cite{howard}, the experimental data suggest that
\begin{equation}
\frac{1}{V} = \frac{1}{V_{max}}\biggl(1 + \frac{K_{m}}{[ATP]}\biggr)
\label{eq-mmenten}
\end{equation}
where $[ATP]$ is the ATP concentration (in mM),
$K_m$ is the Michaelis constant given by $K_{m} = 0.1 $ mM
in this case. $V$ and $V_{max}$ (in ms$^{-1}$) are the
reaction rate and its maximum value respectively.
As mentioned earlier $1/V_{max} \simeq 9$ ms.
Since $1/V = \omega_{h}^{-1} + 5$ ms, we finally get
\begin{equation}
\omega_h^{-1} \simeq \biggl[ 4 + 9 \biggl(
\frac{0.1~\text{mM}}{\text{ATP concentration (in
mM)}}\biggr)
\biggr] \text{ms}
\label{MM}
\end{equation}
so that the allowed biologically relevant range of
$\omega_h$ is $0 \leq \omega_h \leq 250$~s$^{-1}$.

Up to now, experimental investigations could not determine the 
parameter $c$. We therefore treat it as a free parameter in the
following to study the effects that it has on the phase diagram,
position of shocks etc. Comparison with empirical results then
might help to get an estimate for $c$.
%%%%%%%%%%%%%%%%%%%%%%%%%%%%%%%%%%%%%%%%%%%%%%%
%%%%%%%%%%%%%%%%%%%%%%%%%%%%%%%%%%%%%%%%%%%%%%%

%%%%%%%%%%%%%%%%%%%%%%%%%%%%%%%%%%%%%%%%%%%%%%%%%%%%%%%%%%%%%
\subsection{Mean-field equations}
%%%%%%%%%%%%%%%%%%%%%%%%%%%%%%%%%%%%%%%%%%%%%%%%%%%%%%%%%%%%%

Let us denote the probabilities of finding a KIF1A molecule in
the states $1$ and $2$ at the lattice site $i$ at time $t$ by the
symbols $S_i$ and $W_i$, respectively. In mean-field approximation, 
the master equations for the dynamics of the interacting KIF1A 
motors in the bulk of the system are given by
\begin{eqnarray}
\frac{dS_i}{dt}&=&\omega_a (1-S_i-W_i) -\omega_h S_i -\omega_d S_i\nonumber\\
&&+\omega_s W_i +\omega_f W_{i-1}(1-S_i-W_i)\nonumber\\
&& + (1-c)\, \omega_f W_{i}(S_{i+1}+W_{i+1}),\label{eqrc1}\\
\frac{dW_i}{dt}&=&
-(\omega_s + \omega_f) W_i(1-S_{i+1}-W_{i+1}) +\omega_h S_i \nonumber\\
&&-(\omega_s+(1-c)\omega_f) W_i (S_{i+1}+W_{i+1}) \nonumber\\
&&-\omega_b W_i (2-S_{i+1}-W_{i+1}-S_{i-1}-W_{i-1})\nonumber\\
&&+\omega_b (W_{i-1}+W_{i+1})(1-S_i-W_i) \nonumber\\
&=&-(\omega_s + \omega_f) W_i +\omega_h S_i
+c\, \omega_f W_i (S_{i+1}+W_{i+1}) \nonumber\\
&&-\omega_b W_i (2-S_{i+1}-W_{i+1}-S_{i-1}-W_{i-1})\nonumber\\
&&+\omega_b (W_{i-1}+W_{i+1})(1-S_i-W_i). \label{eqhc1}
\end{eqnarray}
The corresponding equations for the left boundary ($i = 1$) are given by
\begin{eqnarray}
\frac{dS_1}{dt}&=& \alpha (1-S_1-W_1) + \omega_s W_1
- \omega_h S_1 - \gamma_1 S_1 \nonumber\\
 &&+(1-c)\omega_f W_1(S_2+W_2),\\
\frac{dW_1}{dt}&=&  \omega_h S_1 - (\omega_s+\omega_f) W_1
+c \omega_f W_1(S_2+W_2) \nonumber\\
&&- \gamma_2 W_1 + \omega_b W_2(1-S_1-W_1) \nonumber\\
&&- \omega_b W_1(1-S_2-W_2), \label{eq-lbound1}
\end{eqnarray}
while those for the right boundary ($i = L$) are given by
\begin{eqnarray}
\frac{dS_L}{dt}&=& \delta (1-S_L-W_L) +\omega_f W_{L-1}(1-S_L-W_L)
\nonumber\\
&&+ \omega_s W_L
- \omega_h S_L  - \beta_1 S_L,\label{eq-rbound0}\\
\frac{dW_L}{dt}&=&  \omega_h S_L - \omega_s W_L
- \beta_2 W_L \nonumber\\
&&+ \omega_b W_{L-1}(1-S_L-W_L) \nonumber\\
&& - \omega_b W_L(1-S_{L-1}-W_{L-1}).
 \label{eq-rbound1}
\end{eqnarray}
In the following we shall determine solutions of this set of equations
for several cases and compare with the corresponding numerical results 
from computer simulations.

%%%%%%%%%%%%%%%%%%%%%%%%%%%%%%%%%%%%%%%%%%%%%%%%%%%%%%%%%%%%%%%%%
\section{Comparison with other models for motor traffic}
%%%%%%%%%%%%%%%%%%%%%%%%%%%%%%%%%%%%%%%%%%%%%%%%%%%%%%%%%%%%%%%%%
\label{sec-compare}

In this section we compare our model with earlier models of molecular 
motor traffic. The first two subsections describe models developed 
for {\it non-interacting} molecular motors whereas in the last 
subsection we collect the main results for the PFF model which has 
been introduced to study collective effects in motor traffic.
A more detailed comparison with models of interacting motors will 
be taken up later in section \ref{intcomp} of this paper.

Chen \cite{chen} developed a model for single-headed kinesins assuming
a {\it power stroke} mechanism. He assumed that each kinesin can
attain three distinct states which were labelled by the symbols $0$,
$1$ and $2$. The kinesin was assumed to be detached from the
microtubule in the state $0$, but bound to microtubule in the other
two states. The states $1$ and $2$ were assumed to differ from each
other by the amount of their tilt in the direction of motion. The
molecule steps ahead by exactly $8$ nm in one cycle consuming one ATP
molecule. This power-stroke model fails to account for several aspects
of experimental data (for example, the distribution of the steps
sizes, including backward steps) on KIF1A and, therefore, will not be
considered further for quantitative comparison.

%%%%%%%%%%%%%%%%%%%%%%%%%%%%%%%%%%%%%%%%%%%%%%%%%%%%%%%%%%%%%%%%%
\subsection{Comparison with Sasaki's Brownian ratchet model}
%%%%%%%%%%%%%%%%%%%%%%%%%%%%%%%%%%%%%%%%%%%%%%%%%%%%%%%%%%%%%%%%%

In contrast to the power-stroke model developed by Chen \cite{chen},
Sasaki \cite{sasaki} quantified the Brownian-ratchet model for a
single KIF1A motor proposed by Okada and Hirokawa \cite{okada1,okada2}.
He used the standard Fokker-Planck approach \cite{julicher,reimann}.
In this formulation, the particle, which represents a kinesin, is
assumed to be subjected to a time-dependent periodic potential
as given in Fig.~\ref{fig-ratgauss}.
The potential switches from one shape $V_1(x)$ to another
shape $V_2(x)$ with rate $\omega_1$ and the reverse switching takes
place at a rate $\omega_2$. One of the shapes of this potential
$V_1(x)$ is taken to be a periodic repetition of a saw-tooth where
each saw-tooth itself is asymmetric. Suppose, the height of the
maximum of each sawtooth is $U$. The shape of the form of the
potential $V_2(x)$ was assumed to be flat, i.e., $V_2(x) = 0$ for
all $x$. Sasaki calculated the average speed $v$ and the diffusion
coefficient $D$ as functions of $U$, $\omega_1$ and $\omega_2$.

One advantage of our model over Sasaki's model is that we do not
make any ad-hoc assumption regarding the shape of the potential as
the potential does not enter explicitly into our formulation.
It is possible to identify $\omega_1$ in Sasaki's model with
$\omega_h$ in our model. The rate constant $\omega_2$ can be 
related to the rates in our model in the following way: 
$\omega_2=\omega_s+\omega_f$ if the preceding site is unoccupied 
and $\omega_2=\omega_s+(1-c)\omega_f$ if it is occupied.

%%%%%%%%%%%%%%%%%%%%%%%%%%%%%%%%%%%%%%%%%%%%%%%%%%%%%%%%%%%%%%%%%
\subsection{Comparison with Fisher-Kolomeisky multi-step chemical kinetic 
model}
%%%%%%%%%%%%%%%%%%%%%%%%%%%%%%%%%%%%%%%%%%%%%%%%%%%%%%%%%%%%%%%%%

\begin{figure}[htb]
\begin{center}
\vspace{0.5cm}
\includegraphics[width=0.45\textwidth]{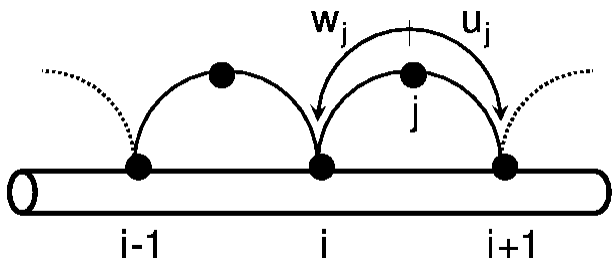}\\
\end{center}
\caption{Fisher-Kolomeisky multi-step chemical kinetic model of 
molecular motors.
}
\label{fig-kolo}
\end{figure}

Next we make a comparison between our model and the multi-step 
chemical kinetic approach developed by Fisher and Kolomeisky 
\cite{kolo1,kolo2,kolo3} for molecular motors. 
In the simplest case of a single filament, the equispaced binding 
sites on a microtubule are assumed to form a one-dimensional lattice. It is 
assumed that there are $M$ distinct discrete intermediate chemical 
states on a biochemical pathway between two consecutive binding sites. 
The motor in state $j_{i}$ (i.e., in chemical state $j$ located 
at spatial position $i$ where $1 \leq j \leq M$, $1 \leq i \leq L$) 
can make transitions to the states $(j+1)_{i}$ and $(j-1)_{i}$ with 
the rates $u_{j}$ and $w_{j}$, respectively (see Fig.~\ref{fig-kolo}). 
Note that we have labelled the chemical states in such a way that 
$M_{i} = 1_{i+1}$ ($M_{i} = 2$ in Fig.~\ref{fig-kolo}) such that, 
completion of the chain in forward (backward) transitions through 
these $M$ states would translocate the motor forward (backward) by 
one lattice spacing. 
 
%%%%%%%%%%%%%%%%%%%%%%%%%%%%%%%%%%%%%%%%%%%%%%%
%%%%%%%%%%%%%%%%%%%%%%%%%%%%%%%%%%%%%%%%%%%%%%%

\begin{figure}[htb]
\begin{center}
\includegraphics[width=0.5\textwidth]{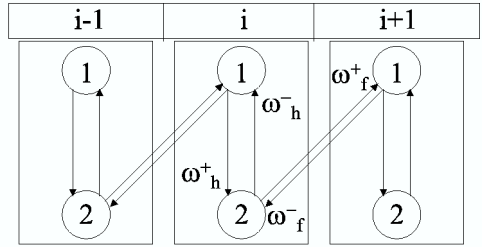}\\
\end{center}
\vspace{-1.5cm}
\caption{In the absence of attachment and detachment, our model is 
equivalent to Fisher-Kolomeisky model shown in Fig.~\ref{fig-kolo}.
}
\label{fig-kolofisher}
\end{figure}

Clearly, in the absence of attachment and detachment of the motors, 
our model for a single KIF1A reduces to the Fisher-Kolomeisky 
multi-step chemical kinetic model of molecular motors on a single 
filament (see Fig.\ref{fig-kolofisher}) where $M = 2$, as 
emphasized by a slight redrawing of our model in Fig.~\ref{fig-kolofisher}. 

%%%%%%%%%%%%%%%%%%%%%%%%%%%%%%%%%%%%%%%%%%%%%%%
%%%%%%%%%%%%%%%%%%%%%%%%%%%%%%%%%%%%%%%%%%%%%%%

Direct quantitative comparison with our model is also 
possible. For example, in the special case where only forward 
transitions are allowed and $M = 2$, the average speed of the 
motor in the Fisher-Kolomeisky model is given by 
\begin{equation}
v = \frac{u_1 u_2}{u_1 + u_2} 
\label{eq-kolo}
\end{equation}
where distance is measured in the units of spacing between 
two successive binding sites ($8$ nm in case of microtubule). 
In Sec.~\ref{sub-nonint} we will derive an analogous expression
for our model, see (\ref{eq-garaij}).

%%%%%%%%%%%%%%%%%%%%%%%%%%%%%%%%%%%%%%%%%%%%%%%%%%%%%%%%%%%%%%%%%%%%%
\subsection{Comparison with PFF-model}
%%%%%%%%%%%%%%%%%%%%%%%%%%%%%%%%%%%%%%%%%%%%%%%%%%%%%%%%%%%%%%%%%%%%%

The Parmeggiani-Franosch-Frey model (PFF model) \cite{frey1} combines
the TASEP with Langmuir kinetics.  The motors are assumed to step
forward one site with rate $p$ if the front site is empty, but do not
move if this site is occupied (exclusion). A backwards movement is not
possible.  In addition, motors can attach to empty sites with rate
$\omega_a$ and detach from a site with rate $\omega_d$. This might be
the simpliest model for intracellular transport including adsorption
and desorption. Although quite basic, it already reproduces the
qualitative behavior of a large class of many-motor systems. It not
only shows high-, low- and maximum-current phases like TASEP, but also
phase coexistence for distinct parameter ranges, while phase domains
are separated by stationary domain walls (shocks). These shocks are
also observed in experiments \cite{nosc}. Shock phases appear if the
Langmuir kinetics are of the same order as motor attachment and
detachment at the ends. It means that in the continious limit where
system size $L\to\infty$, the local attachment and detachment rates
$\omega_a$ and $\omega_d$ have to be rescaled so that the global
attachment and detachment rates defined as $\Omega_a:=\omega_a L,\,\,
\Omega_d:=\omega_d L$ stay constant. One can argue that the topology   
of the phase diagram of the PFF model is quite universal for
systems that, as the PFF model, possess a current-density relation with
one single maximum and the same Langmuir kinetics \cite{popkov}, so 
even more complex models might show similiar qualitative behavior as 
the PFF model.

Although the PFF model reproduces qualitative properties of
intracellular transport quite well, it is difficult to associate the hopping
parameter $p$ quantitatively with experimentally accessible
biochemical quantities because the biochemical processes of a motor
making one step are usually quite complex. The PFF model does not take
into account these processes. Furthermore, it is not possible to
include interactions in the PFF model that only influence particular
transitions of the biochemical states of the motor. The advantage of
our model is the possibility of calibration of the model parameters
with experimentally controllable parameters ATP- or motor protein 
concentration. Through the parameter $c$ we can include, at least
phenomenologically, an interaction that controls the transition from
one state (2) to another (1).

%%%%%%%%%%%%%%%%%%%%%%%%%%%%%%%%%%%%%%%%%%%%%%%%%%%%%%%%%%%%%%%%%%%%
\section{Single motor properties and calibration}
%Discretized model of interacting KIF1A motors
%%%%%%%%%%%%%%%%%%%%%%%%%%%%%%%%%%%%%%%%%%%%%%%%%%%%%%%%%%%%%%%%%%%%

In this section we first investigate the dynamics of our model in the
limit of vanishing inter-motor interactions. This helps us to calibrate
the model properly by comparing with empirical results. Then we compare
the non-interacting limit of our model as well as the corresponding
results with earlier models of non-interacting motors to elucidate
the similarities and differences between them.

%%%%%%%%%%%%%%%%%%%%%%%%%%%%%%%%%%%%%%%%%%%%%%%%%%%%%%%%%%%
\subsection{Calibration of our model in the low-density limit}
%%%%%%%%%%%%%%%%%%%%%%%%%%%%%%%%%%%%%%%%%%%%%%%%%%%%%%%%%%%

An important test of our model would be to check if it reproduces the
single molecule properties in the limit of extremely low density of
the motors. We have already explained earlier how we extracted the
numerical values of the various parameters involved in our model.  The
parameter values $\omega_{a} = \alpha = 1.0 \times 10^{-3}$ s$^{-1}$,
allows realization of the condition of low density of kinesins.  Using
those parameters sets, we carried out computer simulations with
microtubules of fixed length $L=600$ which is the typical number of
binding sites along a microtubule filament. Each run of our simulation
corresponds to a duration of 1 minute of real time if each timestep is
interpreted to to correspond to 1~ms. The numerical results of our
simulations of the model in this limit, including their trend of
variation with the model parameters, are in excellent agreement with
the corresponding experimental results (see Table~\ref{tab-1mol}).

%%%%%%%%%%%%%%%%%%%%%%%%%%%%%%%%%%%%%%%%%%%%%%%%%%%%%%%
\begin{table}
\begin{tabular}{|c|c|c|c|c|} \hline
ATP (mM)&  $\omega_h$ (1/s)& $v$ (nm/ms)&  $D/v$ (nm) & $\tau$ (s)\\\hline
$\infty$ & 250 & 0.201 & 184.8 & 7.22 \\ \hline
0.9      & 200 & 0.176 & 179.1 & 6.94 \\ \hline
0.3375   & 150 & 0.153 & 188.2 & 6.98 \\ \hline
0.15     & 100 & 0.124 & 178.7 & 6.62 \\ \hline
\end{tabular}
\caption{\label{tab-1mol}{Predicted transport properties
in the low-density limit for four different ATP densities.
$\tau$ is calculated by averaging the intervals between attachment
 and detachment of each KIF1A.}}
\end{table}
%%%%%%%%%%%%%%%%%%%%%%%%%%%%%%%%%%%%%%%%%%%%%%%%%%%%%%%

%%%%%%%%%%%%%%%%%%%%%%%%%%%%%%%%%%%%%%%%%%%%%%%%%%%%%%%%%%%%%%%%%%%%%%%%%%%%%%
\subsection{Non-interacting limit of our model: a mean-field analysis}
%%%%%%%%%%%%%%%%%%%%%%%%%%%%%%%%%%%%%%%%%%%%%%%%%%%%%%%%%%%%%%%%%%%%%%%%%%%%%%
\label{sub-nonint}

For the case of a single KIF1A molecule, all interaction terms
can be neglected and the mean-field equations (\ref{eqrc1}), (\ref{eqhc1})
for the bulk dynamics are linearized and simplify to 
\begin{eqnarray}
\frac{dS_i}{dt}&=&\omega_a (1-S_i-W_i) +\omega_f W_{i-1}+ 
\omega_s W_i \nonumber \\
&-& \omega_h S_i -\omega_d S_i,\label{eq-bulk0}\\
\frac{dW_i}{dt}&=&  \omega_h S_i  -\omega_s W_i
-\omega_f W_i \nonumber\\
&+& \omega_b (W_{i-1}+W_{i+1}) - 2 \omega_b W_i  \, .
\label{eq-bulk1}
\end{eqnarray}
The boundary equations (\ref{eq-rbound0})-(\ref{eq-lbound1}) also get 
simplified in a similar way.

Assuming {\em periodic} boundary conditions, the (homogeneous)
solutions $(S_i,W_i)=(S,W)$ of the mean-field equations 
(\ref{eq-bulk0}), (\ref{eq-bulk1}) in the steady-state are found to be
\begin{eqnarray}
S &=& \frac{\omega_a(\omega_s + \omega_f)}{\omega_a(\omega_h + \omega_s 
+ \omega_f) + \omega_d(\omega_s + \omega_f)}\, ,  \label{eq-garais}\\
W &=& \frac{\omega_a \omega_h}{\omega_a(\omega_h + \omega_s + \omega_f) 
+ \omega_d(\omega_s + \omega_f)}\, .
\label{eq-garaiw}
\end{eqnarray}
The corresponding flux is given by
\begin{eqnarray}
J &=& \omega_f W \nonumber \\
  &=& \frac{\omega_a\omega_h\omega_f}{\omega_f(\omega_a + \omega_d) 
+ \omega_a(\omega_s+\omega_h) + \omega_d\omega_s} \nonumber \\
   &=& \frac{\omega_h}{(1+K) + (\Omega_s + \Omega_h) + \Omega_s K}.
\label{eq-garaij}
\end{eqnarray}
where $K = \omega_d/\omega_a$, $\Omega_h = {\omega_h}/{\omega_f}$ and
$\Omega_s = {\omega_s}/{\omega_f}$.

If we make the correspondence $u_1 = \omega_h$, and $ u_2 = \omega_f$
the expression (\ref{eq-garaij}) for the average speed of KIF1A in our
model, in the special case $\omega_s = 0$ (i.e., no reverse
transition) reduces to the Fisher-Kolomeisky result (\ref{eq-kolo}).
A more general version of the non-interacting limit of our 
model is treated in appendix A.

%%%%%%%%%%%%%%%%%%%%%%%%%%%%%%%%%%%%%%%%%%%%%%%%%%%%%%%%%%%%%%%%%%
\section{Collective flow properties}\label{intcomp}
%%%%%%%%%%%%%%%%%%%%%%%%%%%%%%%%%%%%%%%%%%%%%%%%%%%%%%%%%%%%%%%%%%

In the following we will study the effects of interactions
between motors which lead to interesting collective phenomena.

%%%%%%%%%%%%%%%%%%%%%%%%%%%%%%%%%%%%%%%%%%%%%%%%%%%%%%%%%%%%%%%%%%
\subsection{Collective properties for $c=1$}
%%%%%%%%%%%%%%%%%%%%%%%%%%%%%%%%%%%%%%%%%%%%%%%%%%%%%%%%%%%%%%%%%%

We first look at the case $c=1$ originally studied in \cite{nosc}.
In mean-field approximation the master equations (\ref{eqrc1}), (\ref{eqhc1})
for the dynamics of the interacting KIF1A motors in the bulk of the system
are nonlinear.
Note that each term containing $\omega_f$ is now multiplied by the
factor of the form $(1-S_i-W_i)$ which incorporates the effects
of mutual exclusion.

Assuming {\em periodic} boundary conditions, the solutions
$(S_i, W_i)=(S,W)$ of the mean-field equations  (\ref{eqrc1}), (\ref{eqhc1})
in the steady-state for $c=1$ are found to be
\begin{eqnarray}
S&=&\frac{ -\Omega_h - \Omega_s -  (\Omega_s -1)K  +
{\sqrt{D}} }{2 K(1+K)}\, ,\label{eqr1}\\
W&=&\frac{\Omega_h +\Omega_s + (\Omega_s +1)K
-{\sqrt{D}} }{2 K}\label{eqh1}\, ,
\label{eq-swsteady}
\end{eqnarray}
where $K=\omega_d/\omega_a$,
$\Omega_h=\omega_h/\omega_f$, $\Omega_s=\omega_s/\omega_f$, and
\begin{equation}
 D=4\Omega_s K(1+K)+
(\Omega_h +\Omega_s + ( \Omega_s-1)K)^2.
\end{equation}
Thus, the density of the motors, irrespective of the internal
``chemical'' state, attached to the microtubule is given by
\begin{equation}
\rho = S + W \frac{\Omega_h + \Omega_s + (\Omega_s+1)K 
- \sqrt{D} + 2}{2(1+K)}\, .
\end{equation}
This is the analogue of the Langmuir density for this model; it is
determined by the three parameters $K$, $\Omega_h$ and $\Omega_s$.
Note that, as expected on physical grounds,
$S + W \rightarrow 1$ as $K \rightarrow 0$ whereas
$S+W \rightarrow 0$ as $K \rightarrow \infty$.
The probability of finding an empty binding site on a microtubule is $KS$ as
the stationary solution satisfies the equation $S+W+KS=1$.

The steady-state flux of the motors along their microtubule tracks is given by
\begin{equation}
J=\omega_f W(1-S-W).
\label{eq-flux}
\end{equation}
Using the expressions (\ref{eq-swsteady}) for $S$ and $W$ in equation
(\ref{eq-flux}) for the flux we get the analytical expression
\begin{equation}
J = \frac{\omega_f\biggl[K^2 - \biggl(\Omega_h + (1+K) \Omega_s -
\sqrt{D}\biggr)^2\biggr]}{4K(1+K)}.
\label{eq-flux2}
\end{equation}

%%%%%%%%%%%%%%%%%%%%%%%%%%%%%%%%%%%%%%%%%%%%%%%
\begin{figure}[htb]
\begin{center}
\includegraphics[width=0.4\textwidth]{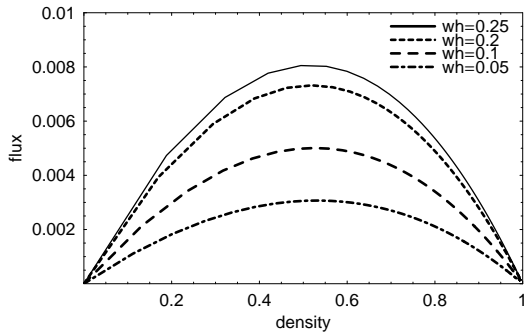}
\end{center}
\caption{Fundamental diagram (i.e., flux-versus-density relation)
for the traffic flow of KIF1A in our model.
}
\label{fig-fdkif1a}
\end{figure}
%%%%%%%%%%%%%%%%%%%%%%%%%%%%%%%%%%%%%%%%%%%%%%%

The flux obtained from the expression (\ref{eq-flux2}) for several
different values of $\omega_h$ are plotted as the fundamental diagrams
for this model in Fig.~\ref{fig-fdkif1a}. Note that, in general, this
model lacks the particle-hole symmetry. This is obvious from the flux
can be recast in general as
\begin{equation}
 J=\frac{\omega_h}{\omega_h+\omega_s+\omega_f(1-c\rho)}\omega_f\rho(1-\rho).
  \label{fundall}
\end{equation}
This is easily derived by substituting the relation $\rho=S+W$ and
the constant solution of (\ref{eqhc1})
\begin{equation}
 -(\omega_s+\omega_f)W+\omega_h(\rho-W)+c\omega_fW\rho=0
\label{eq27}
\end{equation}
into the definition of the flux (\ref{eq-flux}).

Next we consider two limiting cases. In
case I ($\omega_{f} \ll \omega_{h} \simeq \omega_{s}$) the
forward movement is the rate-limiting process and in
case II ($\omega_{h} \ll \omega_{f} \simeq \omega_{s}$) the
availability of ATP and/or rate of hydrolysis is the rate-limiting
process.

%%%%%%%%%%%%%%%%%%%%%%%%%%%%%%%%%%%%%%%%%%%%%%%%
\subsubsection{Case I ($\omega_{f} \ll \omega_{h} \simeq \omega_{s}$)}
%%%%%%%%%%%%%%%%%%%%%%%%%%%%%%%%%%%%%%%%%%%%%%%%
In this case,
\begin{widetext}
\begin{equation}
S \simeq \frac{1}{2K(1+K)}
\left[ - \Omega_s(1+K) + K +\frac{1}{2}(1+K)(3+K)\left(
\Omega_s - K^2\right) - \frac{1}{8}\Omega_s(1+K)^2(3+K)^2 + K^2 -
\frac{K^4}{2\Omega_h}\right]\, ,
\label{eq-rcase1}
\end{equation}
\begin{equation}
W \simeq \frac{1}{2K}\left[ \Omega_s(1+K) + K - \frac{1}{2}(1+K)(3+K)\left(
\Omega_s - K^2\right) + \frac{1}{8}\Omega_s(1+K)^2(3+K)^2 - K^2 +
\frac{K^4}{2\Omega_h}\right] \, ,
\label{eq-hcase1}
\end{equation}
\end{widetext}
so that the total density is
\begin {equation}
\rho = S + W \simeq \frac{2 + \Omega_s(1+K) +K -(\sqrt{D}
- \Omega_h)}{2(1 + K)}\, .
\label{eq-rhocase1}
\end{equation}
Therefore, in this case, the steady-state flux is given by
\begin{equation}
J = \omega_f W (1 - \rho) \simeq \frac{\omega_f\left[\rho(1+K)
- 1\right](1 - \rho)}{K}\, .
\label{eq-jcase1}
\end{equation}
In this case, in addition, if $K = \frac{\omega_d}{\omega_a} \ll 1$,
i.e., detachments are rare compared to attachments, $K^2$ can be
treated as negligibly small and, hence, equations (\ref{eq-rcase1}),
(\ref{eq-hcase1}) and (\ref{eq-rhocase1}) simplify to the forms
\begin{equation}
S \simeq \frac{1}{2(1+K)} \, ,
\end{equation}
\begin{equation}
W \simeq \frac{1}{2} \, ,
\end{equation}
\begin{equation}
\rho \simeq \frac{2+K}{2(1+K)}\, .
\end{equation}
The corresponding formula for the flux becomes
\begin{equation}
J \simeq q_{\rm eff}^{(1)} \rho (1 - \rho)
\end{equation}
where
\begin{equation}
q_{\rm eff}^{(1)} = \frac{\omega_f(1+K)}{2+K} \simeq
\frac{\omega_f}{2}\, .
\label{eq-asephop1}
\end{equation}
Note that this effective hopping probability
is also derived directly from (\ref{fundall}) by
putting $\omega_{f} \ll \omega_{h} \simeq \omega_{s}$.

Thus, the result for the flux in the special case
can be interpreted to be that of a system of ``particles''
hopping from one binding site to the next with the effective hopping
probability $q_{\rm eff}^{(1)}$.

However, if we assume only $\omega_{h} \gg \omega_{f}$, but the
relative magnitudes of $\omega_{h}$ and $\omega_{s}$ remains
arbitrary,
\begin{equation}
S \simeq \frac{-\Omega_h - \Omega_s -(\Omega_s - 1)K + \Omega_h +
\Omega_s(1+K) -K}{2K(1+K)} \simeq 0\, ,
\end{equation}
\begin{equation}
W \simeq \frac{\Omega_h + \Omega_s + (\Omega_s + 1)K - \Omega_h -
(1+K)\Omega_s + K}{2K} \simeq 1\, .
\end{equation}
Physically, this situation arises from the fact that, because of
fast hydrolysis, the motors make practically instantaneous transition
to the weakly bound state but, then, remain stuck in that state for
a long time because of the extremely small rate of forward hopping.

%%%%%%%%%%%%%%%%%%%%%%%%%%%%%%%%%%%%%%%%%%%%%%
\subsubsection{Case II ($\omega_{h} \ll \omega_{f} \simeq \omega_{s}$)}
%%%%%%%%%%%%%%%%%%%%%%%%%%%%%%%%%%%%%%%%%%%%%%

In this case also the flux (\ref{fundall}) can be interpreted to be 
that for a TASEP where the particles hop with the effective effective
hopping probability
\begin{equation}
q_{\rm eff}^{(2)} = \omega_h \biggl[\frac{\omega_f}{\omega_s + \omega_f
(1-\rho)}\biggr] \, .
\label{eq-asephop2}
\end{equation}
that depends on the density $\rho$. The specific form of $q_{\rm eff}$ 
in equation (\ref{eq-asephop2}) is easy to interpret physically. A 
tightly bound motor attains the state 2 with the rate $\omega_{h}$ 
and only a fraction $\frac{\omega_f}{\omega_s + \omega_f (1-\rho)}$
of all the transitions from the state 2 lead to forward hopping of 
the motor.

%%%%%%%%%%%%%%%%%%%%%%%%%%%%%%%%%%%%%%%%%%%%%%%%%%%%%%%%%%%%
\subsection{Collective properties for $c=0$}
%%%%%%%%%%%%%%%%%%%%%%%%%%%%%%%%%%%%%%%%%%%%%%%%%%%%%%%%%%%%
We now consider the case $c=0$ where ADP release by the kinesin is 
independent of  the occupation status of the front site. 
Let us study the stationary state of the mean-field equations
(\ref{eqrc1}), (\ref{eqhc1}) in the case $c=0$. From (\ref{eqhc1})
we get
\begin{equation}
S_i = \frac{\omega_s + \omega_f}{\omega_h}W_i
\end{equation}
by neglecting the terms that represent Brownian motion. Substituting
this into (\ref{eqrc1}) we have
\begin{equation}
 \omega_f H_{i-1}(1-H_i)- \omega_f H_{i}(1-H_{i+1})
-\alpha_d H_i + \alpha_a (1-H_i)=0,
\label{PFF}
\end{equation}
where we put
\begin{eqnarray}
 \frac{\omega_s + \omega_f+\omega_h}{\omega_h}W_i &=& H_i\, , \label{rh}\\
 \frac{\omega_s + \omega_f}{\omega_h}\omega_d &=& \alpha_d\, ,  \label{rd}\\
 \frac{\omega_s + \omega_f+\omega_h}{\omega_h}\omega_a &=& \alpha_a
\, .\label{ra}
\end{eqnarray}
Eq.~(\ref{PFF}) is the same equation as for the stationary PFF model.
Therefore, the phase diagram of this model would be identical
to that of the PFF model in mean field approximation if we rescale
all the parameters by (\ref{rh})-(\ref{ra}). One has to stress that
this model is not \emph{exactly} identical to the PFF model. While mean
field approximation is exact for the PFF model in the continuous
limit, our model shows correlations \cite{pgdiplom} that lead to different 
density profiles and phase diagrams (see Sec.~\ref{sub-contlimit}). 
Nevertheless, the topological structure of the phase diagrams remains
the same in both models and the differences are not quite large.

So far we have discussed two possible scenarios of ADP release  
by kinesin; in one of these the process depends on the status of
occupation of the target site ($c=1$) whereas it is autonomous in the 
other ($c=0$).
To our knowledge, at present, the available experimental data
can not rule out either of these two scenarios of ATP hydrolysis by
kinesins.
Therefore, we have introduced the parameter $c$  that
interpolates both these possible scenarios.
As we have seen in this section, the extended model interpolates,
at least on the level of mean-field theory, between the PFF model
and the model introduced in \cite{nosc}. In the following section
we will discuss some properties of the extended model including
case $0 < c < 1$ in more
detail. We focus on the density profiles and especially the properties
of shocks.

%%%%%%%%%%%%%%%%%%%%%%%%%%%%%%%%%%%%%%%%%%%%%%%%%%%%%%%%%%%%
\section{Position of the shock}
%%%%%%%%%%%%%%%%%%%%%%%%%%%%%%%%%%%%%%%%%%%%%%%%%%%%%%%%%%%%

\begin{figure}[tb]
\begin{center}
\includegraphics[width=0.4\textwidth]{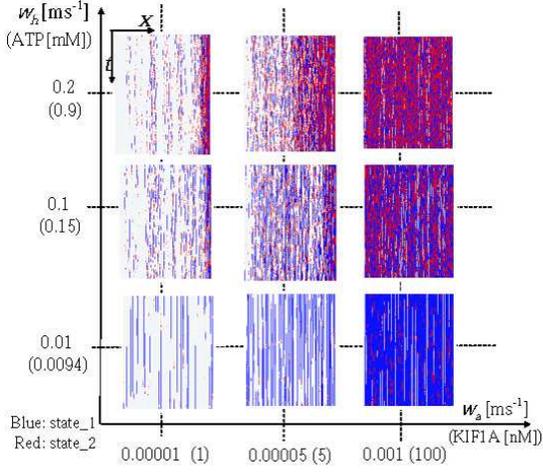}
\end{center}
\caption{(Color online)
 Space-time plot of the model system for $c=1$. Each row of squares 
represents the state of the system at one single instant of time 
whereas successive rows (in the upward direction) correspond to 
the state of the system with increasing time. The blue and red squares 
indicate kinesins in the states $1$ and $2$, respectively, while 
the white squares correspond to empty binding sites on the microtubule. 
Total number of binding sites is $600$, and the configurations of the 
system are displayed for the last $1200$ time steps of a simulation 
run up to a total of $2\times 10^5$ time steps, starting from an 
initial state where all the binding sites on the microtubule were empty.
The other model parameters are $\omega_a=0.3$,$\omega_d=0.2$, 
$\omega_h=400$,$\omega_f=600$,$\omega_s=200$,$\omega_b=50$ for bulk,
and $\alpha=50$,$\beta_1=\beta_2=700$,$\gamma_1=\gamma_2=\delta=0$
for boundaries. 
}
\label{fig-spa}
\end{figure}

One of the interesting results of the model is the existence of a 
domain wall separating 
the high-density and low-density phases in the steady state of the 
system. One such configuration is shown in the space-time diagram 
in Fig.~\ref{fig-spa}.
% and the corresponding time-averaged density 
%profile of the motors are plotted in Fig.~\ref{fig-averagerh}. 
In this section we shall determine the position of the shock, i.e.,
the domain wall, and the trends of its variation with the model 
parameters $\omega_a$ and $\omega_f$, etc. 
%%%%%%%%%%%%%%%%%%%%%%%%%%%%%%%%%%%%%%%%%%%%%%%
%%%%%%%%%%%%%%%%%%%%%%%%%%%%%%%%%%%%%%%%%%%%%%%

%%%%%%%%%%%%%%%%%%%%%%%%%%%%%%%%%%%%%%%%%%%%%%%
%\begin{figure}[htb]
%\begin{center}
%\vspace{0.5cm}
%\includegraphics[width=0.45\textwidth]{greulich9a.eps}
%\includegraphics[width=0.45\textwidth]{greulich9b.eps}
%\end{center}
%\caption{Density profiles of the model KIF1A kinesin motors attached 
%to the microtubule in the (a) state $1$ and (b) state $2$, respectively. 
%All the parameter values are identical to those in Fig.\ref{fig-spa}.
%}
%\label{fig-averagerh}
%\end{figure}
%%%%%%%%%%%%%%%%%%%%%%%%%%%%%%%%%%%%%%%%%%%%%%%

%%%%%%%%%%%%%%%%%%%%%%%%%%%%%%%%%%%%%%%%%%%%%%%%%%%%%%%%%%%%
\subsection{Analytical treatment in the continuum limit}
%%%%%%%%%%%%%%%%%%%%%%%%%%%%%%%%%%%%%%%%%%%%%%%%%%%%%%%%%%%%
\label{sub-contlimit}
Let us first introduce the variable $x = \frac{i-1}{L-1}$; since 
$1 \leq i \leq L$, we have $0 \leq x \leq 1$.
We map our system $(S_i, W_i)$ into $(S(x),W(x))$, and consider
the continuum limit by considering $L$ to be large enough:
\begin{equation}
S(x\pm \epsilon) = S(x) \pm \epsilon \frac{\partial S}{\partial x}
+ \frac{\epsilon^2}{2} \frac{\partial^2 S}{\partial x^2}
\end{equation}
for $S(x \pm \epsilon)$ and a similar expansion for $W(x \pm \epsilon)$,
where $\epsilon=1/L$.
Using this Taylor expansion, we get
\begin{eqnarray}
\frac{\partial S(x,t)}{\partial t} &=&\omega_a (1-S-W)
+ \omega_s W - (\omega_h+\omega_d) S \nonumber \\
&+& \omega_f [W-\epsilon\frac{\partial W(x,t)}{\partial x}]
 (1-S-W)\nonumber \\
 &+&(1-c)\omega_fW(S+W \nonumber\\
 && +\epsilon\frac{\partial S(x,t)}{\partial x}
  +\epsilon\frac{\partial W(x,t)}{\partial x}
  )
\, ,\nonumber \\
\frac{\partial W(x,t)}{\partial t} &=& c\omega_f W
\left[S+W+\frac{\partial S(x,t)}{\partial x}\epsilon +
\frac{\partial W(x,t)}{\partial x}\epsilon\right]\nonumber \\
&-& (\omega_s+\omega_f) W + \omega_h S\, .\
\label{eq-swcont1}
\end{eqnarray}

In the stationary state, we have
\begin{eqnarray}
\epsilon\frac{\partial S(x)}{\partial x} &=&
 \frac{\omega_s+\omega_f}{c\,\omega_f}
 -\frac{\omega_h}{c\,\omega_f} \frac{S}{W} - (S+W) -
\frac{\partial W(x)}{\partial x}\epsilon\, , \nonumber \\
\epsilon\frac{\partial W(x)}{\partial x} &=& \frac{1}{1-S-W}
(\frac{\omega_a}{\omega_f}
+\frac{\omega_s+\omega_f-c\,\omega_a}{c\,\omega_f}W
\nonumber\\
&-&\frac{\omega_h+c(\omega_a+\omega_d)}{c\,\omega_f}S-W(S+W).
\label{eq-swcont3}
\end{eqnarray}
Moreover, from the left boundary equations, by letting $S_1=S_2=S(0)$
and $W_1=W_2=W(0)$, we obtain
\begin{eqnarray}
&&\alpha[1 - S(0) - W(0)] - \omega_h S(0) + \omega_s W(0) \nonumber\\
&&\hspace{1cm}+(1-c) \omega_f W(0)[S(0) + W(0)]=0 \\
&&-(\omega_s+\omega_f)W(0)+\omega_h S(0) \nonumber\\
 &&\hspace{2cm}+c\omega_f W(0)(S(0)+W(0))=0,
\label{eq-lcont1}
\end{eqnarray}
and, hence,
\begin{eqnarray}
S(0) &=& \frac{\alpha-[c\,\alpha(\alpha-\omega_s)/\omega_f]}{c\,\alpha+\omega_h}
\, , \nonumber \\
W(0) &=& \frac{\alpha}{\omega_f}\, .
\label{eq-lcont3}
\end{eqnarray}
Similarly from the right boundary conditions
\begin{eqnarray}
&&-\omega_h S(1) + \omega_s W(1) -\beta S(1)\nonumber\\
&&\hspace{1cm}+\omega_f W(1)[1 - S(1) - W(1)]=0, \\
&&-\omega_s W(1) + \omega_h S(1) -\beta W(1)=0.
\label{eq-rcont2}
\end{eqnarray}
Solving these equations we have
\begin{eqnarray}
S(1) &=& \frac{\omega_s + \beta}{\omega_h}
 \biggl[
\frac{\omega_h}{\omega_h+\omega_s+\beta} - \frac{\beta}{\omega_f}
\biggr]\, ,
\nonumber \\
W(1) &=& \frac{\omega_h}{\omega_h+\omega_s+\beta} - \frac{\beta}{\omega_f}
\, .
\label{eq-rcont3}
\end{eqnarray}

Note that the pair of coupled equations (\ref{eq-swcont3}) involves
only the first order derivatives of $S$ and $W$ with respect to $x$
whereas we have two sets of boundary conditions (\ref{eq-lcont3}) and
(\ref{eq-rcont3}). Therefore, if we integrate the equations
(\ref{eq-swcont3}) using the boundary conditions (\ref{eq-lcont3}),
the solution may not, in general, match smoothly with the other
solution obtained for the same equation using the boundary conditions
(\ref{eq-rcont3}). The discontinuity corresponds to a shock or
domain wall.

The continuity
condition gives $J_l=J_r$ where the flow just at the left side is
denoted by $J_l=\omega_f h_l (1-r_l - h_l)$ and that at the right is $J_r$.
Thus we integrate (\ref{eq-swcont3}) numerically by using
(\ref{eq-lcont3}) to the right end,
and we also integrate them by 
(\ref{eq-rcont3}) to the left end, and seek the point where $J_l=J_r$ is
attained.

\begin{figure}[htb]
\begin{center}
\vspace{0.6cm}
\includegraphics[bb = 45 50 715 530, width=0.45\textwidth]{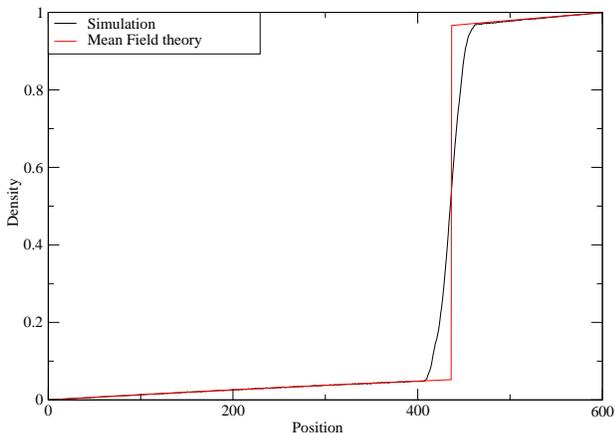}
\end{center}
\caption{(Color online)
 An example of a density profile with shock obtained
 by integrating the mean-field
 equations as well as that of numerical result.
 Parameters are $\omega_h=200, \omega_a=0.01$.
}
\label{denpro}
\end{figure}

We have investigated the shock position by changing
the values of $\omega_a$ and $\omega_h$.
The results are given in Table~\ref{sposi} in the case $c=1$,
which quantitatively agree with numerical simulations as shown in
Fig.~\ref{denpro}.
%%%%%%%%%%%%%%%%%%%%%%%%%%%%%%%%%%%%%%%%%%%%%%%
%%%%%%%%%%%%%%%%%%%%%%%%%%%%%%%%%%%%%%%%%%%%%%%

%%%%%%%%%%%%%%%%%%%%%%%%%%%%%%%%%%%%%%%%%%%%%%%%%%%%%%%%%%
\begin{table}
\begin{tabular}{|c|c|c|c|c|} \hline
$\omega_h$& $\omega_a$=0.01 & $\omega_a$=0.025 & $\omega_a$=0.05 & $\omega_a$=0.065 \\ \hline
200 & 0.725 & 0.5 & 0.318 & 0.253 \\ \hline
150      & 0.776 & 0.571 & 0.382 & 0.311 \\ \hline
125   & 0.808 & 0.618 & 0.425 & 0.35 \\ \hline
\end{tabular}
\caption{The position of the localized shock.
Parameters are $L=600$, $c=1$, $\omega_d=\beta=0.1$, $\alpha=\omega_a$,
 $\omega_f=145$ and $\omega_s=55$.}
\label{sposi}
\end{table}
%%%%%%%%%%%%%%%%%%%%%%%%%%%%%%%%%%%%%%%%%%%%%%%%%%%%%%%%%%%

%%%%%%%%%%%%%%%%%%%%%%%%%%%%%%%%%%%%%%%%%%%%%%%%%%%%%%%%%%%%
\subsection{Shock position from simulations}
%%%%%%%%%%%%%%%%%%%%%%%%%%%%%%%%%%%%%%%%%%%%%%%%%%%%%%%%%%%%

In this subsection we locate the position of the shock in our model 
using a new {\it shock tracking probe} (STP) which is an extension   
of {\it ``second class particles''} (SCP) \cite{janleb} used earlier 
for locating domain walls in computer simulations of driven-diffusive 
lattice gas models defined on a discrete lattice. In the standard TASEP 
model, a SCP is defined as one that behaves as a particle while 
exchanging position with a hole and behaves as a hole while exchanging 
position with a particle. As a result, the second class particle 
has a tendency to get localized at the domain wall (or, the shock). 
Other types of STP have also been considered in the literature 
\cite{sakuntala}

The rules for the movements of the STP in our model of KIF1A traffic 
have been prescribed by extending those for SCP in TASEP. 
Let us use the symbols $\bar 1$ and $\bar 2$ to denote the STPs which 
correspond to the states 1 and 2, respectively, of the particles. Now,  
in the special case $c = 1$, we define the following rules for the 
movements of the STPs:
\begin{equation}
\begin{array}{ll}
\bar 1 \to \bar 2 & \mbox{with rate } \omega_{h} \\
\bar 2 \to \bar 1 & \mbox{with rate } \omega_{s} \\
\bar 2 0 \to 0\bar 1 & \mbox{with rate } \omega_{f} \\
\bar 2 0 \to 0\bar 2 & \mbox{with rate } \omega_{b} \\
0\bar 2 \to \bar 2 0 & \mbox{with rate } \omega_{b} \\
0X\cdots X\bar X \to \bar X X\cdots XX & \mbox{with rate } \omega_{a} \\
%\vspace{3mm} & \\
2\bar X \to \bar X 1 & \mbox{with rate } \omega_{f} \\
2\bar X \to \bar X 2 & \mbox{with rate } \omega_{b} \\
\bar X 2 \to 2 \bar X & \mbox{with rate } \omega_{b} 
\end{array}
\label{eq-fiftyfour}
\end{equation}
with $X$ and $\bar X$ denoting occupation in either state of particles or 
STPs respectively, while $\cdots$ denotes a line of sites occupied by 
particles. Further extension of these rules for arbitrary $c$ is 
straightforward.

These rules satisfy the STP-principle: if the selected site is a STP
it behaves like a particle, while if the selected site is a particle
it treats STPs in its vicinity as holes (by changing sites
respectively). Note that there is no attachment and detachment of
STPs. This is no problem after all, because $\omega_a$ and $\omega_d$
scale like $\frac{1}{L}$ with system size $L$ and we are only looking
at a local quantity (the shock position), so they can be neglected for
large systems (which we are interested in). Besides, for real (finite)
systems they are negligibly small compared to the other rates
$\omega_{h},\omega_{s},\omega_{f}$ and $\omega_{b}$. On the other hand, 
if the STPs were allowed to detach, the undesirable possibility of 
losing all the STPs through detachments could not be ruled out. 
Moreover, allowing STPs to attach and detach like the real particles 
would involve further subtleties of normalization during computation 
of averaged quantities.

A STP, which is not located at the shock, has a tendency to move to the 
shock position. Moreover, if a STP is already located at the shock, it 
follows the shock as the shock moves. For the purpose of illustration, 
consider first an {\em idealized} shock of the form ...0000XXXXXX... 
Inserting a STP in either the low density region or the high density 
region it is obvious from the rules given in (\ref{eq-fiftyfour}) that 
it will, on the average, move in the direction of the shock. However, 
in our model, the observed shocks are not ideal. Instead a few particles 
(holes) will appear in the low (high) density region. As a first 
approximation, one can assume that these particles (holes) are isolated, 
e.g. configurations like ...00X000XXX0XX... Again, by careful use of the 
rules (\ref{eq-fiftyfour}), one can show that the preferred motion of 
the STP is towards the location of the shock also in such realistic 
situations. This argument can be refined even further. In appendix B we 
present an analytical argument in mean-field approximation which supports 
the heuristic arguments used in the illustrative examples in this paragraph.

In addition to the rules listed above we define the following 
{\it fusion rules}:
\begin{equation}
\begin{array}{ll}
\bar 2\bar X \to 0\bar 1 & \mbox{with } \omega_{f} \\
\bar 2\bar X \to 0\bar 2 & \mbox{with } \omega_{b} \\
\bar X\bar 2 \to \bar 2 0 & \mbox{with } \omega_{b} 
\label{eq-fiftyfive}
\end{array}
\end{equation}
The fusion rules ensure that \emph{if} a shock exists, there will be a 
single STP in the system after sufficiently long time. This rule is 
extremely convenient because the lone STP will uniquely define the 
position of the sharp shock rather than a wide region of contiguous 
STPs separating the high-density and low-density regions.  

For the practical implementation of the STPs on the computer, one has 
to select the initial positions of the STPs. We chose to put one STP 
at each end of the system at the beginning of the simulation. If a 
shock can exist in the system, the STPs move to the shock position, 
fuse and, finally, indicate the shock position. We determined the 
shock position in the stationary state by averaging over the fluctuating 
positions of the lone STP in the steady state. In contrast, survival 
of two STPs in the steady state of the system indicates absence of 
any shock; instead, these two STPs indicate the formation of boundary 
layers. Although the latter phenomenon could be interesting, we shall  
not discuss it here. We have compared the shock position obtained 
following the STP approach with that inferred from the density profiles 
measured by computer simulations of our model. These comparisons 
established that the rules (\ref{eq-fiftyfour}) and (\ref{eq-fiftyfive}),  
indeed, yield the correct results.

\begin{figure}[htb]
\begin{center}
\vspace{0.5cm}
\includegraphics[width=0.30\textwidth,angle=-90]{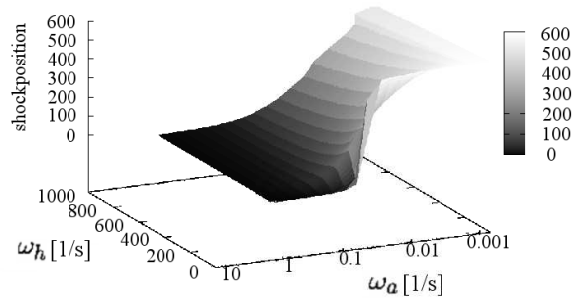}
\end{center}
\vspace{-0.3cm}
\caption{Shock position as function of $\omega_a$ and $\omega_h$
obtained from STP simulations.
}
\label{fig-scpdiag}
\end{figure}

We determined numerically the mean position of shock in a system with
$L=600$ sites as a function of $\omega_a$ and $\omega_h$ which is
shown as a 3D-plot in Fig.~\ref{fig-scpdiag}.

%%%%%%%%%%%%%%%%%%%%%%%%%%%%%%%%%%%%%%%%%%%%%%%
%%%%%%%%%%%%%%%%%%%%%%%%%%%%%%%%%%%%%%%%%%%%%%%

\begin{figure}[htb]
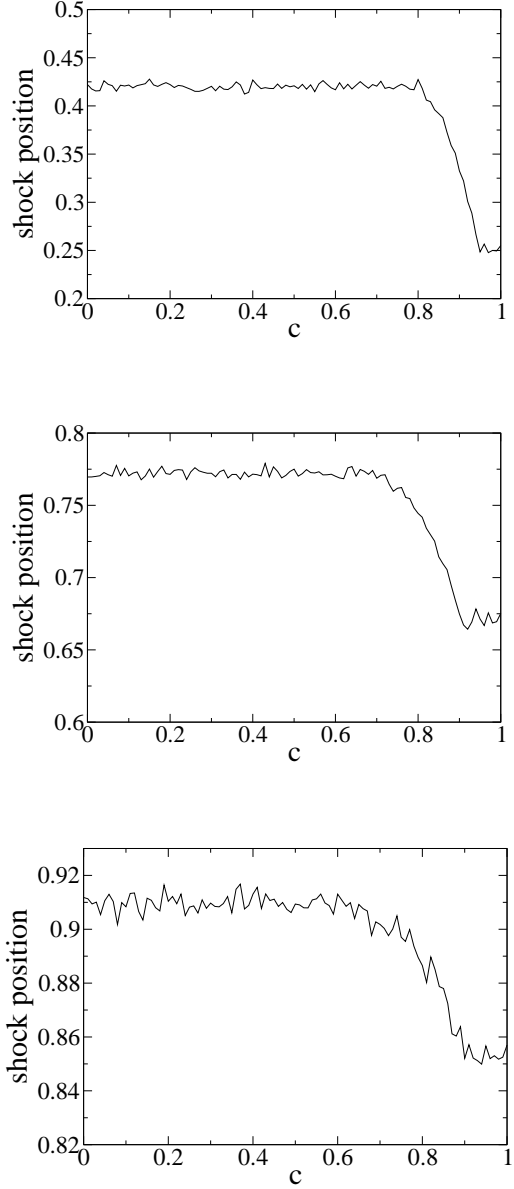

\includegraphics[height=4.5cm]{greulich11a.eps}\\[1.1cm]
\includegraphics[height=4.5cm]{greulich11b.eps}\\[1.1cm]
\includegraphics[height=4.5cm]{greulich11c.eps}\\[0.3cm]
\caption{\label{shkpos(c)} Variation of the shock positions with 
the interaction parameter $c$ for system size $L=3000$, $\omega_d=0.1$ 
and: 
\emph{top}: $\omega_s=55,\,\omega_f=145$, $\omega_a=0.007$, 
$\omega_h=180$; 
\emph{middle}: $\omega_s=100,\,\omega_f=200$, $\omega_a=0.1$, 
$\omega_h=300$; 
\emph{bottom}: $\omega_s=100,\,\omega_f=300$, $\omega_a=0.01$, 
$\omega_h=130$ }
\end{figure}

In Fig.~\ref{shkpos(c)} we have plotted the shock positions as a 
function of the parameter $c$ for different choices of $\omega_s$
and $\omega_f$. In this figure, one observes two plateaus connected by
a decaying domain (leftshift of the shock position). It seems that
upper plateau approximately ends for $c=1-\omega_s-\omega_f$. Further 
detailed investigations will be needed to decide whether the sharp 
change in the position of the shock at this value of $c$ indicates 
merely a crossover or a signature of a genuine phase transition.

%%%%%%%%%%%%%%%%%%%%%%%%%%%%%%%%%%%%%%%%%%%%%%%%%%%%%%%%%%%%%%%%%
%%%%%%%%%%%%%%%%%%%%%%%%%%%%%%%%%%%%%%%%%%%%%%%%%%%%%%%%%%%%%%%%%

%%%%%%%%%%%%%%%%%%%%%%%%%%%%%%%%%%%%%%%%%%%%%%%%%%%%%%%%%%%
\section{Analytical Phase diagram without Langmuir dynamics }
%%%%%%%%%%%%%%%%%%%%%%%%%%%%%%%%%%%%%%%%%%%%%%%%%%%%%%%%%%%

In this section we derive the phase diagram in the plane spanned 
by the boundary rates $\alpha$ and $\beta$ for the special case of 
our model where attachments and detachment of the motors do not take 
place. In other words, we derive the phase diagram of our model in the 
$\alpha-\beta$-plane in the absence of Langmuir kinetics. We use the
domain wall theory proposed in \cite{kolo} to derive this phase diagram 
from the flow-density relation (\ref{fundall}) of the corresponding 
periodic system. From this study, one can calculate the collective 
velocity and the shock velocity which determine the dynamics of the 
density profiles of the open system. Note that, because of the 
translational invariance of the periodic system, $S$ and $W$ show
constant density profiles.

\begin{figure}[htb]
\begin{center}
\vspace{0.5cm}
\includegraphics[width=0.45\textwidth]{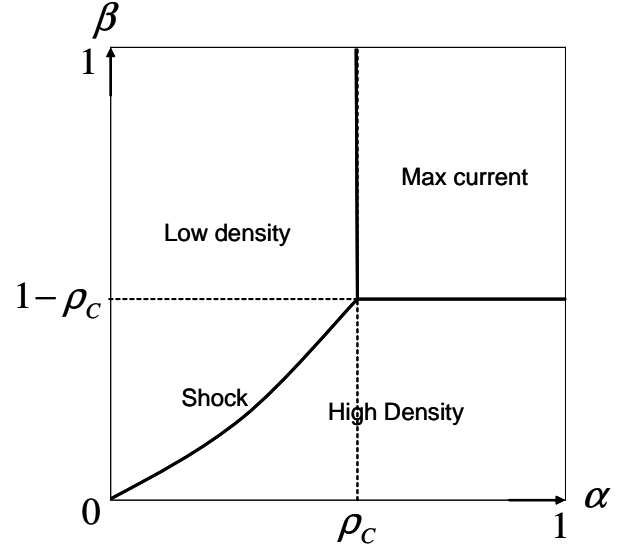}
\end{center}
\vspace{-0.3cm}
\caption{Phase diagram of this system without Langmuir dynamics.
}
\label{phaseab}
\end{figure}

The collective velocity $v_c$ of this system is given by
\begin{eqnarray}
 v_c&=&\frac{\partial J(\rho)}{\partial \rho}\nonumber\\
 &=&\omega_f\omega_h
  \frac{c\omega_f\rho^2-2(\omega_h+\omega_s+\omega_f)\rho+
  \omega_h+\omega_s+\omega_f}{(\omega_h+\omega_s+\omega_f(1-c\rho))^2}.
\nonumber\\
\end{eqnarray}
Thus $v_c=0$ gives the critical density
\begin{equation}
 \rho_c=k-\sqrt{k(k-1)}
\end{equation}
where
\begin{equation}
 k=\frac{\omega_h+\omega_s+\omega_f}{c\omega_f} 
\end{equation}
for the case $c \ne 0$, and $\rho_c=1/2$ for the case $c=0$.
Note that $k$ is always larger than 1.
Next we calculate the shock velocity
\begin{equation}
 S=\frac{J(\rho_L)-J(\rho_R)}{\rho_L-\rho_R},
\end{equation}
where we take $\rho_L=\alpha$ and $\rho_R=1-\beta$. Then we have
\begin{equation}
 S=(\omega_h+\omega_s+\omega_f)(\beta-\alpha)+c\omega_f\alpha(1-\beta).
\end{equation}
From $S=0$ we obtain the first order phase transition curve
\begin{equation}
 \beta=(1-k)\left(1+\frac{k}{\alpha-k}\right),
\end{equation}
that starts at $\alpha=0$ and ends at $\alpha=\rho_c$
(Fig.~\ref{phaseab}).
This curve separates the low and high density phase.
%%%%%%%%%%%%%%%%%%%%%%%%%%%%%%%%%%%%%%%%%%%%%%%
%%%%%%%%%%%%%%%%%%%%%%%%%%%%%%%%%%%%%%%%%%%%%%%

%%%%%%%%%%%%%%%%%%%%%%%%%%%%%%%%%%%%%%%%%%%%%%%%%%%%%%%%%%%%%%%%%%
\section{Experimental investigation with KIF1A}
%%%%%%%%%%%%%%%%%%%%%%%%%%%%%%%%%%%%%%%%%%%%%%%%%%%%%%%%%%%%%%%%%%

%%%%%%%%%%%%%%%%%%%%%%%%%%%%%%%%%%%%%%%%%%%%%%%%%%%%%%%%%%%%%%%%%
%\subsection{Materials and methods} 
%%%%%%%%%%%%%%%%%%%%%%%%%%%%%%%%%%%%%%%%%%%%%%%%%%%%%%%%%%%%%%%%%

\begin{figure}[tb]
\begin{center}
\vspace{0.3cm}
\includegraphics[width=0.45\textwidth]{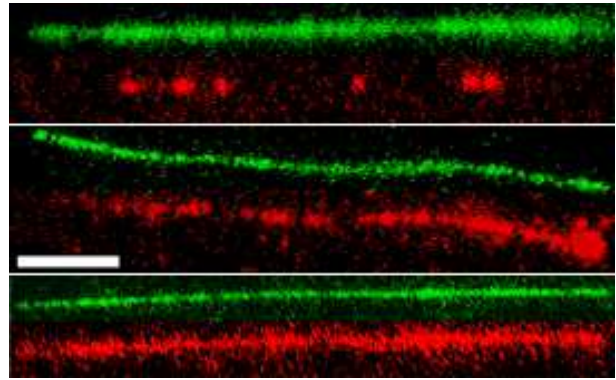}\\
\end{center}
\caption{(Color online)
 A ``comet-like'' structure formed by
 kinesins (red) on the microtubule (green). The high-density
 and low-density regions are clearly separated in this image.
 The white bar length is 2 $\mu$m. 
}
\vspace{-0.3cm}
\label{fig-comet}
\end{figure}

In the experiments performed by Okada \cite{nosc}, microtubules labeled 
with a green fluorescent dye were immobilized on the top surface of the 
cell. The single-headed kinesins labeled with a red fluorescent dye were 
then introduced into the cell together with with ATP. The movement of 
the motor proteins was observed using imaging techniques of optical 
microscopy described in \cite{okada1}.
A ``comet-like'' structure, as shown in Fig.~\ref{fig-comet}, was formed 
by the kinesins (red) on the microtubule (green). The first two images 
from the top, which correspond to low and moderate densities, respectively, 
were taken under essentially same conditions, but the lowermost image in 
the figure was taken with smaller intensifier gain, because it is too 
bright for the intensifier.

No special filtering was applied to the original image.  Each red
fluorescent spot in Fig.~\ref{fig-comet} normally corresponds to a
single fluorescently-labeled kinesin molecule, if the density is not
too high (top panel of Fig.~\ref{fig-comet} is a typical example of
such cases).  Due to the optical resolution limit (about 500 nm), more
than one kinesin can together form a single brighter spot when the
motors are too close to be resolved (as happens, for example, in the
middle panel of Fig.~\ref{fig-comet}). Nevertheless, even in such
situations, the number of fluorochromes in each spot can be estimated
from its intensity. At much higher densities (for example, that
corresponding to the bottom panel of Fig.~\ref{fig-comet}),
fluorescent signals are no longer separable as spots. Even in such
cases, the density of fluorochromes can be estimated from their
intensity profile. However, Okada measured the intensity profile just
to confirm that each spot corresponds to a single kinesin molecule in
the lowest density experiment. In other words, at low densities, the
density of the fluorescent spots gives a good estimate for the density
of kinesins.  But, at higher densities, the spot density gives an
underestimate of the kinesin density due to the overlap of fluorescent
spots (which are not visually separable because of the limited
resolution).

It is true that, under normal physiological conditions, the global 
density of motors in a cell never oversaturates the microtubule surface 
as happened in Okada's experiment described above. However, so far as 
the in-vivo situations are concerned, the motors and microtubules are 
heterogeneously distributed in cells. Thus, the local density of motors 
and microtubule surfaces might be a direct determinant of the formation 
of motor traffic jam within cells during in-vivo experiments. Moreover, 
in pathological situations, traffic jam on microtubule-based transport 
systems, such as axonal transport, is not rare. In fact, such traffic 
jams have been implicated in many neurodegenerative diseases 
\cite{goldd1,goldd2,mandeld1}. Many putative factors may contribute to 
the ``jammorigenesis''; these include the population of the active motor 
proteins, the presence of the inactive motor proteins, the number of 
``obstacles'' on the microtubule surface such as microtubule associated 
proteins, and so on. Obviously, these factors should be, ultimately, 
incorporated into a more ``realistic'' extended version of our model in 
order to explicitly account for the observed ``jammorigenesis''. The 
current version of our model is just the minimal one.

These experimental results have three important implications. First, 
traffic jam can actually take place in living cells at least in
some experimental conditions. Second, the local concentration, rather  
than the global concentration, of the motors determines whether or 
not jam will form in a living cell. Even in the overexpressing cells, 
the overall concentration of motors is much lower than that of tubulin. 
But still ``comet'' is formed. Third, negative regulation systems, 
which are not included in the current version of our model, prevent 
jam formation in physiological situations.

%%%%%%%%%%%%%%%%%%%%%%%%%%%%%%%%%%%%%%%%%%%%%%%
%%%%%%%%%%%%%%%%%%%%%%%%%%%%%%%%%%%%%%%%%%%%%%%

%%%%%%%%%%%%%%%%%%%%%%%%%%%%%%%%%%%%%%%%%%%%%%%%%%%%%%%%%%%%
\section{Discussion}
%%%%%%%%%%%%%%%%%%%%%%%%%%%%%%%%%%%%%%%%%%%%%%%%%%%%%%%%%%%%

In this paper we have proposed a biologically motivated extension of
our recent quantitative model \cite{nosc} describing traffic-like
collective movement of single-headed kinesin motors KIF1A.  
 The dynamics of the system has been formulated in terms of a stochastic
  process where position of a motor is repesented by a discrete
  variable and time is continuous. The model explicitly captures the
  most essential features of the biochemical cycle of each motor by
  assigning two discrete internal (``chemical'' or ``conformational'')
  states to each motor. The model not only takes into account the exclusion
  interactions, as in the previous models, but also includes a possible
  interaction of motors that controls ADP release rates by introducing
  a free parameter $c$. To our knowledge \cite{okadapc}, it is not 
  possible even to establish the existence of this mechano-chemical 
  interaction with the experimental data currently available in the 
  literature. However, we hope that our results reported here will 
  help in developing experimental methods which will not only test the 
  existence of this interaction but also its strength if it exists.
  For example, we have predicted the dependence of the shock position 
  on $c$ (and, therefore, that of $c$ on the shock position). Thus, 
  at least in prinicple, one could determine $c$ by comparing the 
  experimentally measured shock position with this relation. The 
  $c$-dependence of some of the other quantities reported here may 
  provide alternative, and possibly, more direct way of estimating the 
  strength of this mechano-chemical interaction.

We have compared and contrasted our model and the results with earlier
generic models of single motors as well as those of motor traffic. Our 
analytical treatment of the dynamical equations in the continuum limit 
(i.e., a limit in which the spatial position of each motor is denoted 
by a continuous variable) has also established the occurrence of a 
non-propagating shock in this model. We have also calculated the position 
of this shock numerically using the method of second class particles. 

Mean field treatment of the rate equations for $c=0$ showed 
that this special case of our model is equivalent to the simpler PFF 
model which also predicts two-phase coexistence (where the two phases 
are separated by a non-propagating shock). One can argue analytically 
\cite{pgdiplom,popkov}, as we also observed in simulations,  
that the general features of the $\alpha-\beta$-phase diagrams of our 
model is the same as those for the PFF model. Thus, the PFF model, in 
spite of its simplicity, captures the essential generic features of 
intracellular transport. But, it is not possible to make direct 
quantitative comparison between the predictions of the PFF model and 
experimental data as the parameters of the PFF model are not accessible 
to direct biochemical experiments. In contrast, our model captures the 
essential features of the internal biochemical transitions of each 
single-headed kinesin and we could establish a one to one correspondence 
between our model parameters and measurable quantities. The
concentrations of the kinesin motors and ATP are two such important
parameters both of which which are variable \emph{in-vivo} and can be
controlled in \emph{in-vitro}-experiments. We have reported the phase 
diagram of our model in the plane spanned by these experimentally 
accessible parameters.

Finally, we have summarized evidences for the formation of molecular
motor jam from Okada's in-vitro experiments \cite{nosc} and discussed
their relevance in intra-cellular transport under physiological
conditions.

%%%%%%%%%%%%%%%%%%%%%%%%%%%%%%%%%%%%%%%%%%%%%%%%%%%%%%%%%%%%%
\begin{acknowledgments}
We thank Yasushi Okada for the experimental results which we
already reported in our earlier joint letter \cite{nosc} as
well as for many discussions, comments and suggestions.
One of the authors (DC) thanks Joe Howard and Frank J\"ulicher for 
their constructive criticism of our work and the Max-Planck Institute 
for Physics of Complex Systems, Dresden, for hospitality during a 
visit when part of this manuscript was prepared. DC also acknowledges 
partial support of this work by a research grant from the Council of
Scientific and Industrial Research (CSIR) of the government of India.\\
\end{acknowledgments}
%%%%%%%%%%%%%%%%%%%%%%%%%%%%%%%%%%%%%%%%%%%%%%%%%%%%%%%%%%%%%

%%%%%%%%%%%%%%%%%%%%%%%%%%%%%%%%%%%%%%%%%%%%%%%%%%%%%%%%%%%%%%%%%
%\appendix
%\section{A generalized model of non-interacting motors}
\noindent{{\bf Appendix A:} A GENERALIZED MODEL OF NON-INTERACTING MOTORS }\\
%%%%%%%%%%%%%%%%%%%%%%%%%%%%%%%%%%%%%%%%%%%%%%%%%%%%%%%%%%%%%%%%%

In order to make a comparison between the non-interacting limit of
our model and the earlier models of non-interacting molecular
motors, we consider here a slightly more general model which
allows ``reverse'' transitions for each of the ``forward'' transitions.
Then we show that the non-interacting limit of our model is a
special case of the general model while some other special cases
correspond to earlier models of non-interacting motors.
                                                                               
%%%%%%%%%%%%%%%%%%%%%%%%%%%%%%%%%%%%%%%%%%%%%%%
\begin{figure}[htb]
\begin{center}
\includegraphics[angle=-90,width=0.5\textwidth]{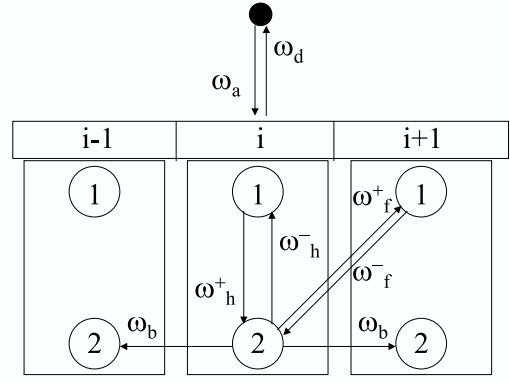}
\end{center}
\vspace{-1cm}
\caption{Schematic description of a general three-state model of
a single molecular motor.
}
\label{fig-kif1agen}
\end{figure}
%%%%%%%%%%%%%%%%%%%%%%%%%%%%%%%%%%%%%%%%%%%%%%%

Consider the multi-step chemical kinetic scheme shown in the
Fig.~\ref{fig-kif1agen}. Note that this generalized scheme
\cite{garaithesis} allows a transition from the strongly bound 
state at $i+1$ to the weakly bound state at $i$ with the rate constant
$\omega_{f}^{-}$ which is not allowed in our model shown in
fig.\ref{fig-kifbrat}. In fact, in this generalized scheme,
corresponding to every forward step (those corresponding to
$\omega_{f}^{+}$, $\omega_{h}^{+}$ and $\omega_{b}$) there is
a backward step (corresponding to $\omega_{f}^{-}$,
$\omega_{h}^{-}$ and $\omega_{b}$, respectively).  This 
generalization is in the spirit of Fisher-Kolomeisky-type 
multi-step chemical kinetic models of molecular motors 
\cite{kolo1,kolo2,kolo3} where each of the reactions are 
allowed to be reversible, albeit with different rate constants, 
in general.

In the mean-field limit the the master equations governing the dynamics
of this general model in the bulk are given by
\begin{eqnarray}
\frac{dS_i}{dt} &=& \omega_a(1-S_i-W_i) + \omega_f^+W_{i-1} \nonumber \\
&+& \omega_h^-W_i - \omega_h^+S_i - \omega_f^-S_i - \omega_dS_i,  
\end{eqnarray}
\begin{eqnarray}
\frac{dW_i}{dt} &=& \omega_h^+S_i + \omega_f^-S_{i+1}
- \omega_h^-W_i - \omega_f^+W_i  \nonumber \\
&+& \omega_b(W_{i-1} + W_{i+1})- 2\omega_b W_i.
\end{eqnarray}
Imposing periodic boundary conditions, the steady state solutions for 
$S$ and $W$ can be written as 
\begin{eqnarray}
S = \frac{\omega_a(\omega_h^- + \omega_f^+)}{\omega_a(\omega_h^- 
+\omega_f^+ + \omega_h^+ + \omega_f^-) + \omega_d(\omega_h^- + \omega_f^+)},  
\label{eq-gens}
\end{eqnarray}
\begin{eqnarray}
W = \frac{\omega_a(\omega_h^+ + \omega_f^-)}{\omega_a(\omega_h^- 
+ \omega_f^+ + \omega_h^+ + \omega_f^-) + \omega_d(\omega_h^- + \omega_f^+)}
\, . 
\label{eq-genw}
\end{eqnarray}
Hence,
\begin{eqnarray}
S+W = \frac{\omega_a(\omega_h^- + \omega_f^+ + \omega_h^+ 
+ \omega_f^-)}{\omega_a(\omega_h^- + \omega_f^+ + \omega_h^+ 
+ \omega_f^-) + \omega_d(\omega_h^- + \omega_f^+)}.\, 
\label{eq-genspw}
\end{eqnarray}
The corresponding steady-state flux 
\begin{eqnarray}
J = W_{i} \omega_{f}^{+} - S_{i+1} \omega_{f}^{-}\,
\end{eqnarray}
is given by 
\begin{eqnarray}
J = \frac{\omega_f^+\omega_h^+ - \omega_f^-\omega_h^-}{(K+1)(\omega_h^- 
+ \omega_f^+) + (\omega_h^+ + \omega_f^-)}.
\label{eq-genj}
\end{eqnarray}

The relation between this generalized model of non-interacting 
motors and the non-interacting limit of our model is quite 
straightforward.
In the special case $\omega_f^- = 0$, using the identification 
$\omega_f^+ = \omega_f$, $\omega_h^+ = \omega_h$ and 
$\omega_{h}^{-} = \omega_{s}$, the equations (\ref{eq-gens}), 
(\ref{eq-genw}) and (\ref{eq-genj}) reduce to the equations 
(\ref{eq-garais}), (\ref{eq-garaiw}) and (\ref{eq-garaij}), 
respectively.\\

%%%%%%%%%%%%%%%%%%%%%%%%%%%%%%%%%%%%%%%%%%%%%%%%%%%%%%%%%%%%%%%%%
%\appendix
%\section{A MF argument for movement of STP and shock position}
\noindent{{\bf Appendix B:} A MF ARGUMENT FOR MOVEMENT OF STP AND SHOCK POSITION}\\
%%%%%%%%%%%%%%%%%%%%%%%%%%%%%%%%%%%%%%%%%%%%%%%%%%%%%%%%%%%%%%%%%

In this appendix we argue that a STP will move to the location of the 
shock, if a shock exists in the system. Our arguments are based on 
an analysis in the mean-field approximation. The master equations for 
the probabilities of the STP, which correspond to the equations 
(\ref{eqhc1}) for the real particles, are given by
\begin{eqnarray}
\label{scp1}
\frac{d}{dt}S_i^{(2)} &=& \omega_f W_{i-1}^{(2)}
(1-\rho_i^{(1)})+\omega_f W_i^{(1)}S_{i+1}^{(2)}
-\omega_f W_{i-1}^{(1)}S_i^{(2)} \nonumber \\ 
&-& \omega_h S_i^{(2)}+\omega_s W_i^{(2)}
+\omega_f(1-c)W_i^{(2)}\rho_{i+1}^{(1)}, \\
\label{scp2}
\frac{d}{dt}W_i^{(2)} &=& \omega_h S_i^{(2)}
-\omega_s W_i^{(2)}-\omega_f W_i^{(2)}(1-\rho_{i+1}^{(1)}) \nonumber \\ 
&-& \omega_f W_{i-1}^{(1)}W_i^{(2)}
+\omega_f W_i^{(1)}W_{i+1}^{(2)}\nonumber\\
 &&-\omega_f(1-c)W_i^{(2)}\rho_{i+1}^{(1)},
\end{eqnarray}
where $S_i^{(2)}$ and $W_i^{(2)}$ represent the probabilities of finding 
the STP in the weakly ($W^{(2)}$) and strongly ($S^{(2)}$) bound states, 
respectively, at the site $i$; note that $S_i^{(1)}$ and $W_i^{(1)}$ are 
the corresponding probabilities for the real particles. Obviously, 
$\rho_i^{(1,2)}=S_i^{(1,2)}+W_i^{(1,2)}$.

Adding the two equations (\ref{scp1}) and (\ref{scp2}) we obtain
\begin{eqnarray}
\frac{d}{dt}\rho_i^{(2)}&=&(\omega_f W_{i-1}^{(2)}(1-\rho_i^{(1)})-\omega_f W_{i-1}^{(1)}\rho_i^{(2)}) \nonumber \\
&-&(\omega_f W_i^{(2)}(1-\rho_{i+1}^{(1)})-\omega_f W_i^{(1)}\rho_{i+1}^{(2)}).
\end{eqnarray}
Comparing this with the equation of continuity $d\rho_i/dt =J_{i-1}-J_{i}$, 
we identify the current $J_i^{STP}$ of STP on site $i$ to be 
\begin{equation}
\label{scp current}
J_i^{STP}=\omega_f W_i^{(2)}(1-\rho_{i+1}^{(1)})-\omega_f W_i^{(1)}\rho_{i+1}^{(2)}.
\end{equation}

Consider a situation where we have \emph{one} STP in a continuous region 
of particles (with no shock inside), so we can put
 $\rho_i^{(1)}\approx\rho_{i+1}^{(1)}=:\rho^{(1)};\, S_i^{(1)}\approx S_{i+1}^{(1)}=:S^{(1)}$ and $W_i^{(1)}\approx
 W_{i+1}^{(1)}=:W^{(1)}$. 
We assume that, after sufficiently long time, the internal states of the 
STP relax to a stationary state so that the probabilities of finding the 
STP in the strongly-bound and weakly-bound states are independent of time. 
However, the mean position of the STP might still change with time.

Then, 
${\cal {S}}:=\sum_i S_i^{(2)}$ 
is the probability of finding the STP in a strongly bound state, while the 
corresponding probability of finding the STP in the weakly bound state is 
${\cal {W}}:=\sum_i W_i^{(2)}$ where the summations are over an interval 
of length $l$ that contains no shocks and one single STP. Obviously, 
${\cal {S}}+{\cal {W}}=1$. Using (\ref{scp2}) we have 
\begin{eqnarray}
0=\frac{d}{dt}{\cal W} &=& \omega_h{\cal S}-\omega_s{\cal W}
 -\omega_f{\cal W} +\omega_f\rho^{(1)}{\cal W}\nonumber \\
&-& \omega_f(1-c)\rho^{(1)}{\cal W},
\label{eqnpg}
\end{eqnarray}
where we have used the fact that 
\begin{eqnarray}
&\omega_f& \sum_i W_i^{(1)}W_{i+1}^{(2)}-W_{i-1}^{(1)}W_i^{(2)} \nonumber \\
=&\omega_f& W^{(1)}(\sum_i W_{i+1}-\sum_i W_{i-1})=0.
\end{eqnarray}
Solving equation (\ref{eqnpg}) for ${\cal W}$, we obtain
\begin{equation}
{\cal W}=\frac{\omega_h}{\omega_h+\omega_s+(1-c\rho^{(1)})\omega_f}.
\label{Wcal}
\end{equation}

%Because of translational invariance, $\rho_i^{(2)}=\frac{1}{l}$ and, hence,
%\begin{equation}
%\label{W(2)_rho(2)}
%W_i^{(2)}=\frac{{\cal W}}{l}=
%\frac{\omega_h\rho_i^{(2)}}{\omega_h+\omega_s+(1-c\rho^{(1)})\omega_f}
%\end{equation}
%The analogous solution for $W^{(1)}$ is 
%\begin{equation}
%\label{W(1)_rho(1)}
%W^{(1)}=
%\frac{\omega_h\omega_f\rho^{(1)}}{\omega_h+\omega_s+(1-c\rho^{(1)})\omega_f}
%\end{equation}

The results derived above are valid for any density distribution of STPs 
as long as there is a shock-free neighbourhood of the STP and the particles 
are in a steady state. Now consider a specific configuration where a STP is 
given to be located at the site $i$ while its internal state remains 
unspecified. In this case,   
\begin{equation}
\rho_k^{(2)}=\delta_{ik}.
\end{equation}
Then we have ${\cal{W}}=W_i^{(2)}$ for any summation interval that 
includes the site $i$. Of course, $W_k^{(2)} = 0$ for $k \neq i$ for this 
distribution of $\rho_k$. Therefore, using (\ref{Wcal}) we obtain 
%Using (\ref{Wcal}) for this distribution of $\rho_k$ we obtain
\begin{equation}
\label{W(2)_rho(2)}
W_k^{(2)}=\frac{\omega_h\delta_{ik}}{\omega_h+\omega_s+
(1-c\rho^{(1)})\omega_f}.
\end{equation}
The analogous solution for $W^{(1)}$ obtained from (\ref{eq27}) is
\begin{equation}
\label{W(1)_rho(1)}
W^{(1)}=\frac{\omega_h\rho^{(1)}}{\omega_h+
\omega_s+(1-c\rho^{(1)})\omega_f}.
\end{equation}

For the density distribution considered here, we can take the current as 
an effective hopping rate of the STP to the right, i.e., 
$q_i^r=J_i(\lbrace\rho_k^{(2)}=\delta_{ik}\rbrace)$. Similarly, we have 
effective hopping rate of the STP to the left 
$q_i^l=-J_{i-1}(\lbrace\rho_k^{(2)}=\delta_{ik}\rbrace)$.
Inserting (\ref{W(1)_rho(1)}) and (\ref{W(2)_rho(2)}) into 
(\ref{scp current}) for $\rho_k^{(2)}=\delta_{ik}$, we obtain 
\begin{equation}
q_i^r-q_i^l=\frac{\omega_f\omega_h}{\omega_h+\omega_s+(1-c\rho^{(1)})\omega_f}(1-2\rho^{(1)}).
\label{qdif}
\end{equation}
Note that the fraction in (\ref{qdif}) is always positive. Therefore, if 
a STP is in a low density region with $\rho^{(1)}<\frac{1}{2}$, we have 
$q_i^r-q_i^l>0$ and the STP tends to hop to the right. But, if the STP is 
in a high density region with $\rho^{(1)}>\frac{1}{2}$, we have 
$q_i^r-q_i^l<0$ and its prefered direction of hopping is left. Thus, in 
the continuum limit, if there is one shock separating a low density region 
at the left and a high density region at the right, any single STP will be 
driven to this domain wall. For sufficiently long time the average position 
of the STP will be equal to the shock position.

%%%%%%%%%%%%%%%%%%%%%%%%%%%%%%%%%%%%%%%%%%%%%%%%%%%%%%%%%%%%
%%%%%%%%%%%%%%%%%%%%%%%%%%%%%%%%%%%%%%%%%%%%%%%%%%%%%%%%%%%%%%%%%%%%%%%%
%   References
%%%%%%%%%%%%%%%%%%%%%%%%%%%%%%%%%%%%%%%%%%%%%%%%%%%%%%%%%%%%%%%%%%%%%%%%

%\end{references}
%%%%%%%%%%%%%%%%%%%%%%%%%%%%%%%%%%%%%%%%%%%%%%%%%%%%%%%%%%%%%
\end{document}